\documentclass[lettersize,journal]{IEEEtran}
\usepackage{url}
\usepackage[utf8]{inputenc}
\usepackage[small]{caption}
\usepackage{graphicx}
\usepackage{booktabs}
\usepackage{algorithm}
\usepackage{cite}
\usepackage{algorithmic}
\usepackage{multirow}
\usepackage{bm}

\usepackage{booktabs}
\usepackage{algorithm,algorithmic}
\usepackage{amsmath,amsfonts}
\usepackage{algorithmic}
\usepackage{algorithm}
\usepackage{array}
\usepackage[caption=false,font=normalsize,labelfont=sf,textfont=sf]{subfig}
\usepackage{textcomp}
\usepackage{mathtools}
\usepackage{stfloats}
\usepackage{url}
\usepackage{verbatim}
\usepackage{graphicx}
\usepackage{cite}
\usepackage{graphicx}
\usepackage{amsmath}
\usepackage{amssymb}
\usepackage{booktabs}
\usepackage{booktabs}
\usepackage{cite}
\usepackage{url}
\usepackage{bm}
\usepackage{multirow}
\usepackage{booktabs}
\usepackage{algorithm,algorithmic}
\usepackage{footnote}
\usepackage{amsmath,amssymb, amsthm,amsfonts}
\usepackage{xcolor}
\usepackage{hyperref}  
\usepackage{bm}
\usepackage{amssymb}
\usepackage{latexsym}
\usepackage{dsfont}
\usepackage{multirow}
\usepackage{amsfonts}
\usepackage{pifont}
\usepackage{amsmath}
\usepackage{multirow}
\usepackage{graphicx}
\usepackage{array}
\usepackage{amssymb}
\usepackage{amsmath}
\usepackage{algorithmic}
\usepackage{color}
\usepackage{cite}

\usepackage[utf8]{inputenc}
\begin{document}
\title{Hierarchical Local-Global Transformer for Temporal Sentence Grounding}

\author{Xiang Fang, Daizong Liu,~\IEEEmembership{Student~Member,~IEEE}, Pan Zhou,~\IEEEmembership{Senior~Member,~IEEE}, Zichuan Xu,~\IEEEmembership{Senior~Member,~IEEE}, Ruixuan Li,~\IEEEmembership{Member,~IEEE}
\IEEEcompsocitemizethanks{
\IEEEcompsocthanksitem The first two authors contributed equally to the paper writing and experiments. (Corresponding author: Pan Zhou)
\IEEEcompsocthanksitem Xiang Fang is with the Hubei Engineering Research Center on Big Data Security, School of Cyber Science and Engineering Huazhong University of Science and Technology, Wuhan 430074, China (E-mail: xfang9508@gmail.com).
\IEEEcompsocthanksitem Daizong Liu is with Wangxuan Institute of Computer fTechnology, Peking University, No. 128, Zhongguancun North Street, Beijing 100080, China (E-mail: dzliu@stu.pku.edu.cn).
\IEEEcompsocthanksitem Pan Zhou is with the Hubei Engineering Research Center on Big Data Security, School of Cyber Science and Engineering, Huazhong University of Science and Technology, Wuhan 430074, China (E-mail: panzhou@hust.edu.cn).
\IEEEcompsocthanksitem Zichuan Xu is with the school of software, Dalian University of Technology, Dalian 116024 China (E-mail: z.xu@dlut.edu.cn).
\IEEEcompsocthanksitem Ruixuan Li is with the School of Computer Science, and Technology, Huazhong University of Science, and Technology, Wuhan 430074, China (E-mail: rxli@hust.edu.cn).
}}

\maketitle

\begin{abstract}
This paper studies the multimedia problem of temporal sentence grounding (TSG), which aims to accurately determine the specific video segment in an untrimmed video according to a given sentence query. Traditional TSG methods mainly follow the top-down or bottom-up framework and are not end-to-end. They severely rely on time-consuming post-processing to refine the grounding results. Recently, some  transformer-based approaches are proposed to efficiently and effectively model the fine-grained semantic alignment between video and query. Although these methods achieve significant performance to some extent, they equally take frames of the video and words of the query as transformer input for correlating, failing to capture their different levels of granularity with distinct semantics. To address this issue, in this paper, we propose a novel \textbf{H}ierarchical \textbf{L}ocal-\textbf{G}lobal \textbf{T}ransformer (HLGT) to leverage this hierarchy information and model the interactions between different levels of granularity and different modalities for learning more fine-grained multi-modal representations. Specifically, we first split the video and query into individual clips and phrases to learn their local context (adjacent dependency) and global correlation (long-range dependency) via a temporal transformer. Then, a global-local transformer is introduced to learn the interactions between the local-level and global-level semantics for better multi-modal reasoning. Besides, we develop a new cross-modal cycle-consistency loss to enforce interaction between two modalities and encourage the semantic alignment between them. Finally, we design a brand-new cross-modal parallel transformer decoder to integrate the encoded visual and textual features for final grounding.  Extensive experiments on three challenging datasets (ActivityNet Captions, Charades-STA and TACoS) show that our proposed HLGT achieves a new state-of-the-art performance, demonstrating its effectiveness and computational efficiency.
\end{abstract}

\maketitle

\section{Introduction}
\begin{figure}[t]
\centering
\includegraphics[width=0.5\textwidth]{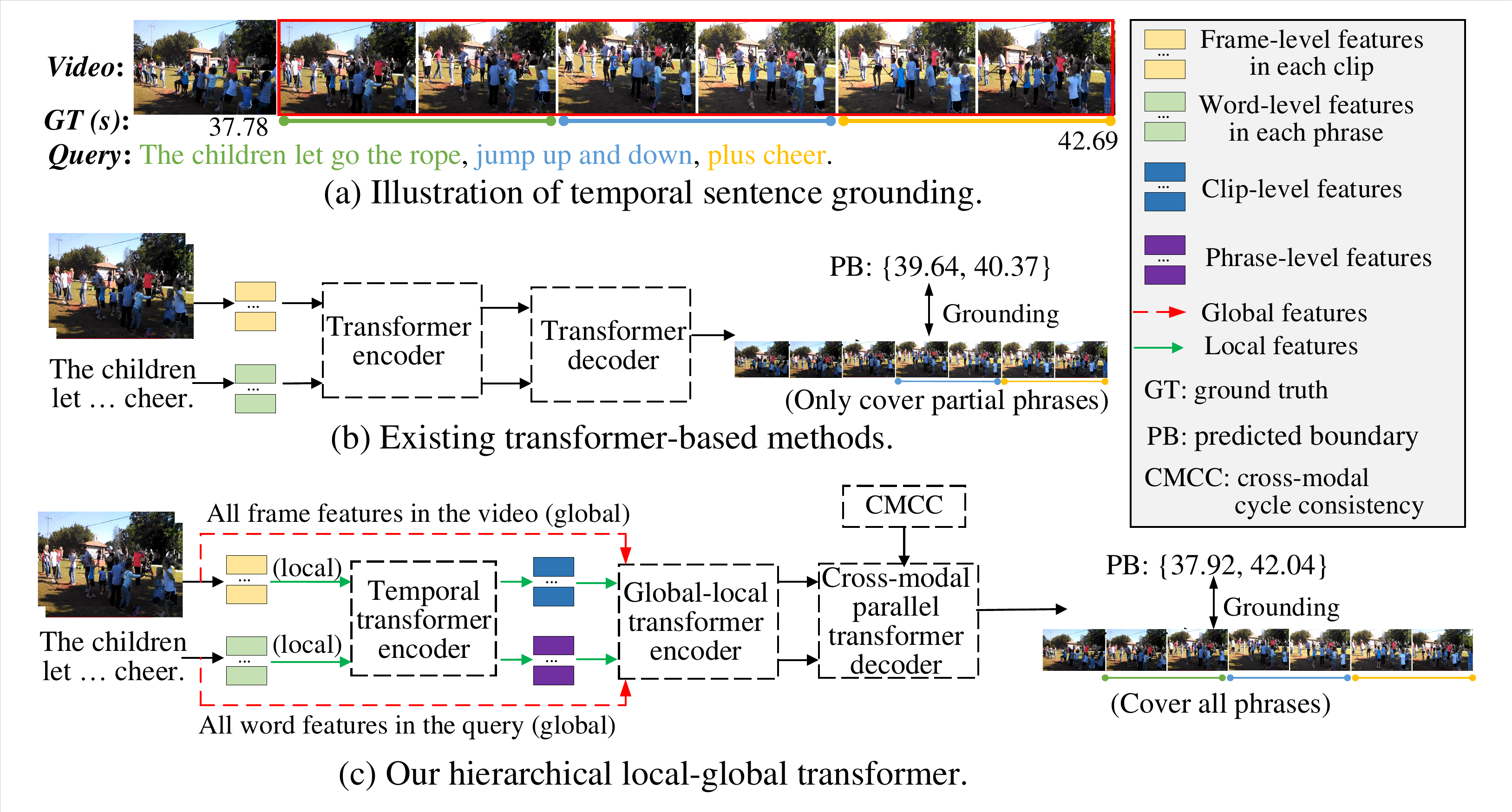}
\caption{(a) An illustrated example of temporal sentence grounding (TSG) where the video contains multiple parts semantically related to different phrases in a sentence. (b) Previous transformer-based methods equally interact frame- and word-level features for semantic alignment, which fail to capture both clip- and phrase-level semantics and are easy to get stuck into the partial semantic matching problem. (c) Our method develops a hierarchical local-global transformer to capture different local clip-phrase semantics and correlate them for globally reasoning the complete sentence.}
\label{fig:intro}
\end{figure}

Temporal sentence grounding (TSG) is a fundamental but important task in multimedia understanding \cite{wang2022cross,zhang2020temporal}. As shown in Fig. \ref{fig:intro}(a), given an untrimmed video, this task aims to predict a specific segment containing the activity related to the semantics of a sentence query. Traditional TSG approaches can be divided into two categories:
1) Top-down approaches \cite{anne2017localizing,chen2018temporally,zhang2019cross,yuan2019semantic,zhang2019learning,liu2022memory}: These methods first pre-define multiple segment proposals and align them with the query for cross-modal semantic matching. The best proposal with the highest similarity score is selected as the predicted segment.
2) Bottom-up approaches \cite{chenrethinking,yuan2019find,mun2020local,zhang2020span,liu2022unsupervised}: These methods directly regress the start and end boundary frames of the target segment or predict boundary probabilities frame-wisely. The predicted segment is obtained through post-processing steps that group or aggregate all frame-wise predictions.
Although the above two types of works have achieved significant performances, they are not end-to-end, and still suffer from the redundant proposal generation/matching process (top-down) and complex post-processing steps (bottom-up) to refine the grounding results.

Recently, transformer-based approaches \cite{cao2021pursuit,zhang2021multi,woo2022explore} are newly introduced to handle the TSG task in an end-to-end manner.
Different from the top-down and bottom-up approaches, they capture more fine-grained interaction between the video-query input and directly output the segment predictions via the effective transformer encoder-decoder architecture \cite{vaswani2017attention,yao2022dual,yao2022wave,li2022contextual,ma2022visualizing} without using any time-consuming pre- and post-processing operation. The general transformer-based pipeline is shown in Fig.~\ref{fig:intro}(b).
It first simply feeds the video frames and query words into the transformer to equally align the semantics between each frame-word pair. Then, the transformer decoder with a direct set prediction \cite{carion2020end,wang2021exploring,li2021referring} is utilized to predict a few learnable segment candidates with corresponding confidence scores.
Thanks to such a simple pipeline and the multi-modal relationship modeling capabilities in a transformer, these transformer-based approaches are both effective and computationally efficient.

However, we argue that existing transformer-based approaches are limited by the bottleneck in capturing the complete correspondence between visual contents and sentence semantics.
Since previous transformer encoders equally interact with each frame-frame/frame-word/word-word pair, they fail to explore both query-related low-level entities and long-range dependent high-level contexts during the semantic alignment.
Thus, they are easy to get stuck into the local minima of partial semantic matching problems, \textit{e.g.}, activating most discriminative parts instead of the full event extend or only partially matching a few semantics in the sentence query.
For instance, as shown in Fig.~\ref{fig:intro}(b), without capturing the individual semantics of different phrases and correlating them for complete reasoning, existing  transformer-based TSG methods only focus on the most discriminative semantics ``jump up and down" and lead to the partial semantic grounding. Considering many real-world video-query pairs involve different levels of granularity, such as frames and words or clips and phrases with distinct semantics, it is crucial to first capture the local query context for modeling the corresponding visual activity and then comprehend the global sentence semantics by correlating all local activities.

To this end, in this paper, we propose a novel Hierarchical Local-Global Transformer (HLGT) to leverage this hierarchy information and model the interactions between different levels of granularity and different modalities for learning more fine-grained cross-modal representations. As shown in Fig.~\ref{fig:intro} (c), we first split the video and query into individual frames and words to compose and learn their clip- and phrase-related dependency via a local temporal transformer encoder. A global temporal transformer encoder is also utilized to correlate the local semantics for complete sentence understanding. Then, we develop a global-local transformer encoder to further learn the complicated interactions between the extracted local- and global-level semantics. Besides, we  develop a new cross-modal cycle consistency loss in the transformer decoder to enforce better cross-modal semantic alignment. Finally, we design a cross-modal parallel transformer decoder to  integrate visual and textual features in parallel to reduce the computational cost. By capturing both local and global granularities in the multi-modal information, our HLGT can capture the complete query semantics for more accurate video grounding.

In summary, the main contributions of our works are:
\begin{itemize}
    \item We present a novel Hierarchical Local-Global Transformer (HLGT), which captures different levels of granularity in both video and query domains to reason the complete semantics for fine-grained grounding. To the best of our knowledge, it is the first time that a multi-level interaction network is proposed to alleviate the limitations of existing transformer-based TSG methods.
    \item We design a cross-modal parallel transformer decoder with a brand-new cross-modal cycle-consistency loss to encourage semantic alignment between visual and language features in the joint embedding space.
    \item We conduct extensive experiments on three challenging benchmarks (ActivityNet Captions, TACoS and Charades-STA) where our proposed HLGT outperforms the state-of-the-arts with clear margins, demonstrating its effectiveness and computational efficiency.
\end{itemize}

\section{Related Work}
\subsection{Traditional TSG Methods}
As a new multimedia task introduced recently \cite{gao2017tall,liu2023exploring,wang2025taylor,fang2026towardsicml,kuai2026dynamic,wang2025point,fang2025your,zhang2025monoattack,fang2020double,liu2024towards,yang2025eood,fang2022multi,fang2026cogniVerse,lei2025exploring,fang2023you,wang2025dypolyseg,fang2025hierarchical,yan2026fit,fang2025adaptive,wang2026topadapter,cai2025imperceptible,fang2026slap,wang2026reasoning,fang2026immuno,wang2026biologically,fang2026disentangling,wang2025reducing,fang2026advancing,fang2026unveiling,wang2026from,liu2023conditional,liu2026attacking,fang2026rethinking,wang2025seeing,fang2026towards,fang2025multi,fang2024fewer,liu2024pandora,fang2024multi,fang2025turing,fang2024not,liu2023hypotheses,fang2024rethinking,liu2024unsupervised,fang2023annotations,xiong2024rethinking,fang2021unbalanced,wang2025prototype,zhang2025manipulating,fang2026align,tang2024reparameterization,fang2025adaptivetai,tang2025simplification,fang2021animc,cai2026towards,fang2020v}, temporal sentence grounding (TSG) aims to identify the start and end timestamps of the most relevant video segment from  an untrimmed video with a  sentence query.
Most works \cite{anne2017localizing,chen2018temporally,zhang2019cross,yuan2019semantic,zhang2019learning,qu2020fine,liu2022memory} have been proposed within a \textit{top-down} framework, which first samples candidate segment proposals from the untrimmed video, then integrates the sentence query with these segments individually, and finally matches them with the query. Although these methods achieve good performances in some cases, they are severely proposal-dependent and time-consuming, which limits their applications.

Although these top-down methods achieve good performances, they are severely proposal-dependent and time-consuming. Without using proposals, the latest methods  \cite{chenrethinking,yuan2019find,mun2020local,zhang2020span,liu2022unsupervised} have been proposed within a \textit{bottom-up} framework, which directly regresses the start and end timestamps of the target segment after interacting the whole video with the query.

Although the above two types of works have achieved significant performances, they are not end-to-end. These methods might suffer from the redundant proposal generation/matching process (top-down) and complex post-processing steps (bottom-up) to refine the grounding results.

\subsection{Transformer-Based TSG Methods}
Recently, some transformer-based TSG approaches \cite{cao2021pursuit,zhang2021multi,woo2022explore} are newly proposed  in an end-to-end manner \cite{zhang2018cross,ging2020coot,he2021end}.
Different from the top-down and bottom-up approaches, they capture more fine-grained interaction between the video-query input and directly output the segment predictions via the effective transformer encoder-decoder architecture \cite{vaswani2017attention} without time-consuming pre- and post-processing operations.
These transformer-based methods first simply feed the video frames and query words into the transformer to equally align the semantics between each frame-word pair. Then, the transformer decoder with a direct set prediction \cite{carion2020end} is used to predict a few learnable segment candidates with corresponding confidence scores. Based on a simple pipeline and the multi-modal relationship modeling capabilities in a transformer, this approach is more effective than traditional methods.

Since many  video-query pairs involve different levels of granularity (\textit{e.g.}, frame-word pairs and clip-phrase pairs), it is crucial to first capture the local query context for modeling the corresponding visual activity and then comprehend the global sentence semantics by correlating all local activities.

\subsection{Cycle-Consistent Learning}
 By utilizing transitivity as the training objective, cycle-consistent learning aims to explore task correlations to regularize training \cite{wu2022blind}, which is widely used in various multimedia fields, such as vision-language navigation~\cite{wang2022counterfactual,chen2022boosting}, text-to-image synthesis \cite{wang2021cycle,song2021enhancing}, and image-text matching \cite{liu2019cyclematch,zhang2022unified}. For example, based on the assumption of cyclical structure, Wang et al.~\cite{wang2019learning} learn visual supervision by tracking forward and backward. In the visual question answering task, Shah et al.~\cite{shah2019cycle} enforce consistency between the generated and the original question by using cycle-consistency for temporal video alignment.
Although these methods address multi-modal tasks, they perform cycle-consistent learning only in the visual domain, which limits their performance. Thus, they cannot apply to our complex TSG task. In a broader sense, our study is the first attempt that explores cycle-consistent learning to the TSG task.
\begin{figure*}[t]
\centering
\includegraphics[width=\textwidth]{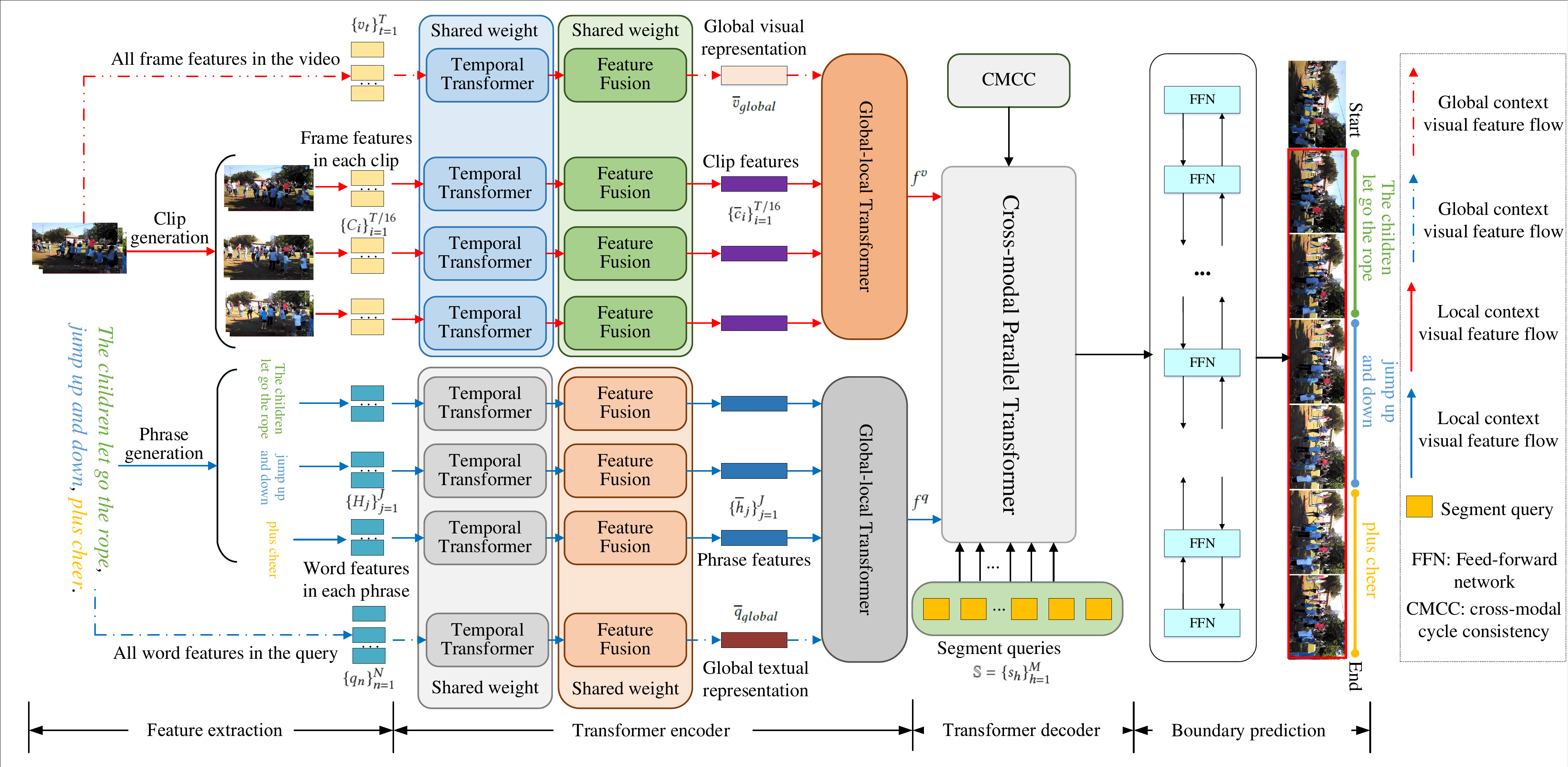}
\caption{The overall pipeline of our proposed HLGT model. Given a video and a query, we first extract frame- and word-level features via the feature extractors. Then, for each modality, we generate local-level features (clip/phrase features) and global-level features via temporal transformer and feature fusion, and integrate these features based on a global-local transformer. After that, we interact the visual and textual features through a cross-modal parallel transformer, where a new cross-modal cycle consistency (CMCC) loss is designed to assist the cross-modal interaction. Finally, the boundary prediction head with feed-back networks is utilized to predict final temporal segments for accurate grounding.
Best view with colors.}
\label{fig:pipeline}
\end{figure*}

\section{The Proposed HLGT Network}
\subsection{Overview}
As a significant multimedia task,  temporal sentence grounding (TSG) aims to localize the precise boundary $(\tau_s, \tau_e)$ of a specific segment from an untrimmed video $\mathcal{V}=\{v_t'\}_{t=1}^T$ semantically corresponding to  a given query $\mathcal{Q}=\{q_n'\}_{n=1}^N$, where $q_n'$ denotes the $n$-th word, $N$ denotes the word number, $\tau_s$ and $\tau_e$ denote the start and end timestamps of the specific segment, $v_t'$ denotes the $t$-th frame, $T$ denotes the frame number, respectively. Recently, some transformer-based approaches \cite{cao2021pursuit,zhang2021multi,woo2022explore} have shown their strong performance to handle the TSG task via the effective transformer encoder-decoder architecture in an end-to-end manner. However, they still suffer from the vanilla transformer design and fail to explore different levels of granularity with distinct but fine-grained semantics in both video and query. Therefore, how to effectively capture and integrate these multi-level cross-modal contexts for better grounding is an emerging issue.

In this section, we present a novel Hierarchical Local-Global Transformer (HLGT), which leverages this hierarchy information and model the interactions between different levels of granularity and multiple modalities for learning more fine-grained multi-modal representations.
As shown in Fig. \ref{fig:pipeline}, the proposed HLGT model consists of four parts, including the multi-modal feature extractors, multi-level transformer encoder, cycle-consistent transformer decoder, and the boundary prediction head.
Given the paired video-query input, we first split the video/query into the clips/phrases, and extract their internal frame- and word-level features via the multi-modal feature extractors.
Then, we capture the relationship between frame/word features within each clip/phrase based on a temporal transformer to integrate the local clip-/phrase-level features.
Meanwhile, we also feed the whole frames/words into another temporal transformer to encode the corresponding global representations.
We fuse the local and global visual and textual tokens by two global-local  transformers to learn the contextualized individual modal semantics.
After that, we introduce a cross-modal parallel transformer decoder to interact the video and query features for semantic alignment in parallel.
Specifically, we develop a new cross-modal cycle consistency (CMCC) loss to assist the multi-modal interaction.
Then, the boundary prediction head is utilized to predict final temporal segments based on the interacted multi-modal representations. Finally, we present the details of each module.
\subsection{Feature Extraction}
\noindent \textbf{Video extractor.}
For video encoding, we first sample every 16 consecutive frames as a clip.
Then, we use a pre-trained Resnet-152 network \cite{he2016deep} to extract the frame-level visual features in each clip.
We denote the extracted video features as $V=\{v_t\}_{t=1}^{T}=\{C_i\}_{i=1}^{T/16} \in \mathbb{R}^{T \times D}$, where $T$ denotes the frame number in the total video, $v_t\in \mathbb{R}^{1 \times D}$ denotes the $t$-th frame, $C_i\in \mathbb{R}^{16 \times D}$ denotes the $i$-th clip and $D$ denotes the visual feature dimension.

\noindent \textbf{Query extractor.}
For query encoding, we first utilize the Glove embedding \cite{PenningtonSM14} to generate the word-level features. The extracted query features are denoted as $Q=\{q_n\}_{n=1}^{N}=\{H_j\}_{j=1}^{J} \in \mathbb{R}^{N \times D}$, where $N$ denotes the word number in the whole query, $q_n\in \mathbb{R}^{1 \times D}$ denotes the $n$-th word feature in the query, $J$ denotes the phrase number, and $D$ denotes the textual feature dimension that is the same as the visual feature dimension in the video extractor. For the $j$-th phrase, $H_j=\{q_k^j\}_{k=1}^{K_j}\in \mathbb{R}^{K_j \times D}$, where $q_k^j$ denotes the $k$-th word in the phrase and $K_j$ denotes the word number in the $j$-th phrase.
Then,  we
follow \cite{liu2020jointly} to split the given query into multiple phrases.  The detailed splitting approach is as follows: to discover the potential phrase-level features, we apply 1D convolutions on the word-level features with different window sizes. At each word location, we compute the inner product of the word feature vectors with convolution filters of three kinds of window sizes, which captures three different-scale phrase features. To maintain the sequence length after convolution process, we zero-pad the sequence vectors when convolution window size is larger than one. The output of the $n$-th word location with window size $s\in\{1,2,3\}$ is formulated by $q_{n,s}^p=tanh(Conv1d(q^w_{n:n+s-1}))\in\mathbb{R}^{1\times D}$, where Conv1d($\cdot$) operates on the windowed features with $D$ kernels. $q_{n,s}^p$ is the phrase-level feature corresponding to $n$-th word location with window size $k$. To find the most contributed phrase at each word location, we then apply max-pooling to obtain the final phrase-level feature $H_j=\{q_k^j\}_{k=1}^{K_j} \in \mathbb{R}^{ K_j \times D}$ by $q_k^j=\max (q_{k,1}^j, q_{k,2}^j, q_{k,3}^j), k\in \{1,2,..., K_j \}$. Thus, we can split the query into multiple phrases.

For the extracted visual features, we focus on two-level frame features in the latter reasoning: frame features in each clip and all the frame features in the whole video. Similarly, we also focus on two-level word features in the latter reasoning: word features in each phrase and all the word features in the given query.
\subsection{Transformer Encoder}
For the transformer encoder, we first feed the extracted frame/word features within each clip/phrase to a temporal transformer followed by a feature fusion module, which can fuse and generate corresponding clip-level/phrase-level features. Then, since these clip-level/phrase-level features can only learn local semantic information of the whole video/query, we also feed all the frame/word features within the whole video/query to the temporal transformer followed by a feature fusion module for learning the global semantic information. Finally, for each modality, we integrate both the global information and the local information by proposing a global-local transformer to generate more contextual features.

\begin{figure}[t]
\centering
\includegraphics[width=0.48\textwidth]{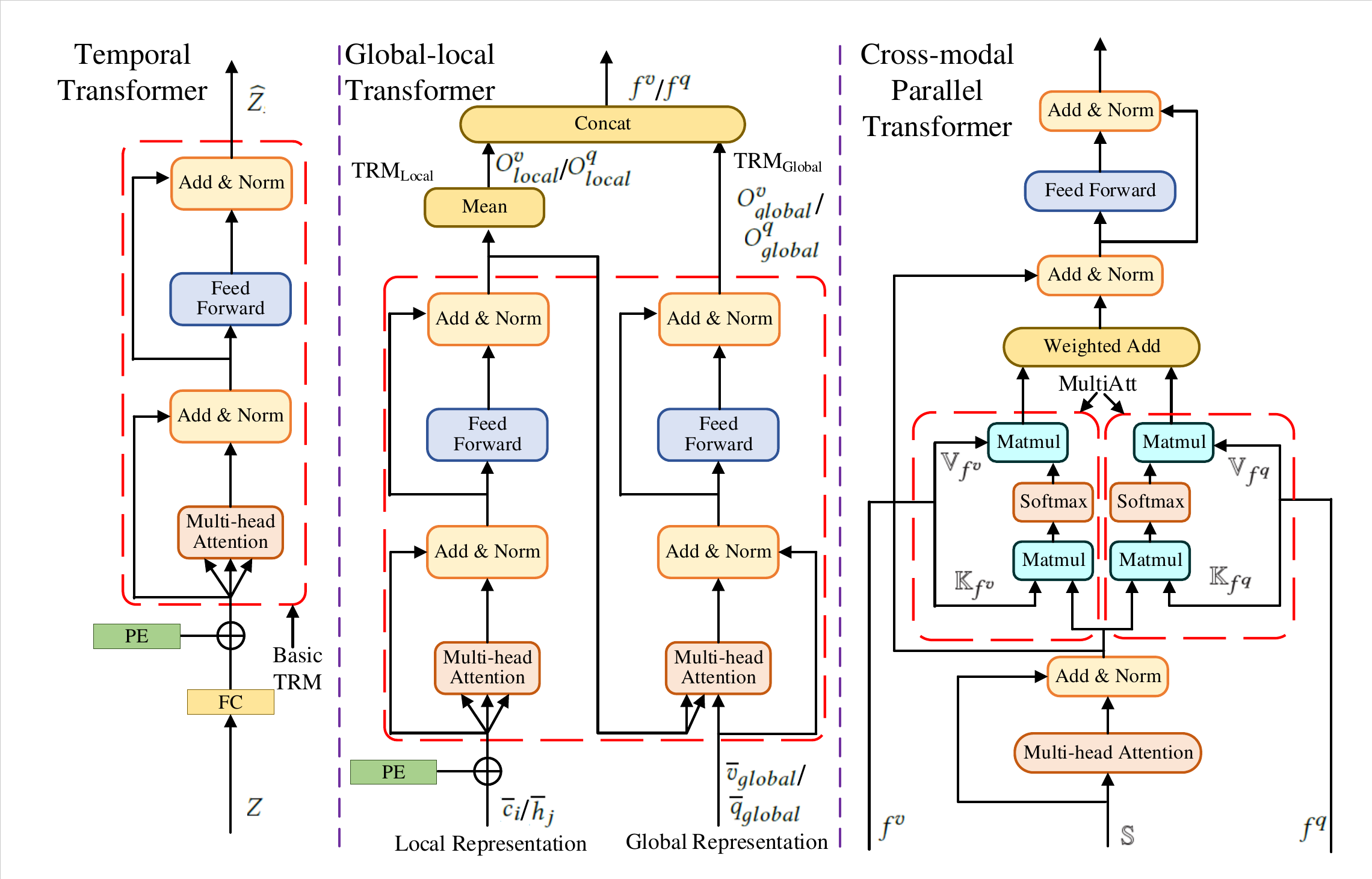}
\caption{The details of our designed transformers.}
\label{fig:transformer}
\end{figure}

\noindent \textbf{Temporal transformer.}
As shown in Fig. \ref{fig:pipeline}, given the extracted frame/word representations within each clip/phrase, we introduce a temporal transformer network with standard attention-blocks to learn the correlations between frames/words for latter clip/phrase-level fusion. Fig. \ref{fig:transformer} shows the details of the temporal transformer.

For ease of description, we first introduce the notation of a standard transformer (called TRM). Considering that transformer architecture contains multi-head self-attention blocks for multi-inputs correlating and updating, we define TRM as:
\begin{align}
\text{TRM}(\mathbb{Q},\mathbb{K},\mathbb{V})=\text{FFN}(\text{Softmax}(\frac{\mathbb{Q}\mathbb{K}^T}{\sqrt{D}}\mathbb{V})),
\label{trm_eq}
\end{align}
where FFN is the feed forward network; $D$ is the feature dimension of the multi-head block; $\mathbb{Q}$, $\mathbb{K}$ and $\mathbb{V}$ are the individual query, key and value in TRM respectively, and they are computed by:
\begin{align}
\mathbb{Q}=W_1Z,\mathbb{K}=W_2Z,\mathbb{V}=W_3Z,
\label{dingyi_kvq}
\end{align}
where $W_1$, $W_2$ and $W_3$ are the embedding weight matrices; $Z$ denotes any sequence (\textit{e.g.}, $C_i$, $H_j$, $Q$ and $V$) as the input of TRM.

Then, in each modality, all the temporal transformers share the same weight.
In the visual branch, if the input of the temporal transformer is the frame-level clip feature $C_i$, and we denote its corresponding output as $\widehat{C}_i$, which is obtained by:
\begin{align}
\widehat{C}_i =\text{TRM}(\mathbb{Q}_{C_i},\mathbb{K}_{C_i},\mathbb{V}_{C_i}),
    \label{v_jian}
\end{align}
where $\mathbb{Q}_{C_i}$, $\mathbb{K}_{C_i}$ and $\mathbb{V}_{C_i}$ are the  query, key and value in the clip-based temporal transformer respectively. To obtain the frame-level video representation $\widehat{V}$, we employ the same temporal transformer on the whole video, and its output is $\widehat{V} =\text{TRM}(\mathbb{Q}_{V},\mathbb{K}_{V},\mathbb{V}_V)$.

Similarly, in the textual branch, if the input of the temporal transformer is $P_j$, its output is $\widehat{H}_j =\text{TRM}(\mathbb{Q}_{{H}_j},\mathbb{K}_{{H}_j},\mathbb{V}_{{H}_j})$, which is the word-level phrase representation.  Likewise, the word-level query representation is $\widehat{Q} =\text{TRM}(\mathbb{Q}_{Q},\mathbb{K}_Q,\mathbb{V}_Q)$.

In the TSG task, both video and query naturally represent at different levels of granularity: a video/query is composed of several clips/phrases, and each clip/phrase contains multiple frames/words. These frames/words in each clip/phrase only contain partial visual/textual information of the specific clip/phrase, which is only the local representation for the given video/query. To obtain the global representation of the given video/query, we also feed all the frame/word features to the temporal transformer to obtain the global visual feature $\overline{v}_{global}$ and the global textual feature $\overline{q}_{global}$.

\noindent \textbf{Feature fusion.} To integrate these frame/word features within each clip/phrase, we introduce a feature fusion module based on an attention-aware approach. For any sequence $Z$, the corresponding attention matrix $A$ is calculated by:
\begin{align}
A = \textup{Softmax}(W_4\textup{Gelu}(W_5Z^T+b_1)+b_2)^T,
    \label{equation_softmax}
\end{align}
where $W_4$ and $W_5$ are two learnable transformation weight matrices; $b_1$ and $b_2$ are two biases; Gelu is the activation function GELU \cite{hendrycks2016gaussian}. Each element in $A$ denotes the pixel-wise attention score whether the frame/word contents contribute to the clip-/phrase- (or video-/query) level semantics. It helps to highlight the foregrounds and filter out the backgrounds for better representation learning.
Thus, based on matrix $A$, we can obtain the fused feature $\overline{z}$ by:
\begin{align}
\overline{z}=\sum_{s=1}^{S}A_s\odot \widehat{Z}_s,
    \label{z_heng}
\end{align}
where $\odot$ is element-wise multiplication; $s$ is the frame/ word number in the clip/phrase (or video/query); $\widehat{Z}$ is the output of temporal transformer with the input of $Z$; $A_s$ and $\widehat{Z}_s$ are the $s$-th attention column of $A$ and $\widehat{Z}$, respectively.

In the visual branch, for the $i$-th clip,  $\widehat{C}_i=\{\widehat{v}_t\}_{t=16(i-1)+1}^{16i}$  is the input of the attention-aware feature fusion module.
Based on Eq.~\eqref{equation_softmax} and \eqref{z_heng}, we fuse all the frame features in $\widehat{C}_i$ into the clip-level feature $\overline{c}_i$. Similarly, we also fuse all the frame features in the whole video into the global visual feature $\overline{v}_{global}$.

For the textual branch, we obtain phrase-level features by fusing the word-level features in the phrase. For instance, we fuse all the word feature in the $j$-th phrase $\widehat{H}_j$ into the phrase-level feature $\overline{h}_j$ based on Eq. \eqref{z_heng}. Besides, we obtain the global textual feature $\overline{q}_{global}$ by fusing all the word features in the given query.

\noindent \textbf{Global-local transformer.} In each modality, the above generated global features (video- or query-level) and local features (clip- or phrase-level) are at different levels. To effectively integrate these cross-level features for more fine-grained feature fusion, we propose a global-local transformer shown in Fig. \ref{fig:transformer}. Specifically, the global-local transformer contains two modules: local transformer $TRM_{Local}$ and global transformer $TRM_{global}$. Both the visual branch and the textual branch have the same global-local transformer architecture. For each modality, the local transformer $TRM_{Local}$ is used to learn the short-term interactions between low-level semantics (adjacent dependency between clips/phrases). The global transformer $TRM_{global}$ aims to model the long-term interactions between local and global representations (global dependency between clip-video or phrase-query).

For $TRM_{Local}$, its core components include: a multi-head attention, a feed-forward layer and a normalization layer. Following \cite{liu2021adaptive}, we append the positional embedding (PE) by using sine function $\sin(\cdot)$ and cosine function $\cos(\cdot)$ of different frequencies:
\begin{align}
    PE_t[2j]&=\sin(\frac{t}{10000^{2j/\varsigma}}),\\ PE_t[2j+1]&=\cos(\frac{t}{10000^{2j/\varsigma}}),
\end{align}
where $2j$ and $2j+1$ are the even and odd indices of the positional embedding; $PE_t$ denotes the positional embedding of the $t$-th position, and $\varsigma$ is the dimension of $PE_t$.
Therefore, the output of the local transformer is:
\begin{align}
O_{local}^v=\frac{16}{T}\sum_{i=1}^{T/16}\text{TRM}_{Local}(\mathbb{Q}_{\overline{c}_i},\mathbb{K}_{\overline{c}_i},\mathbb{V}_{\overline{c}_i}),\\
O_{local}^q=\frac{1}{J}\sum_j^J\text{TRM}_{Local}(\mathbb{Q}_{\overline{h}_j},\mathbb{K}_{\overline{h}_j},\mathbb{V}_{\overline{h}_j}),
\label{out_local}
\end{align}
where $O_{local}^v$ is the visual output of the local transformer and $O_{local}^q$ is the textual output of the local transformer.

The keys/values in $TRM_{Global}$ are from the output of the normalization layer in $TRM_{Local}$, the query is the matrix from the global representation, and we feed both global representation and local representation as input to the multi-head attention block to learn the cross-level correlating and updating. As a result, the TRM block in the global transformer ($TRM_{Global}$) generates attention features to the global representation conditioned on the local representation.
We set $\overline{c}_{local}^i=\text{TRM}_{Local}(\mathbb{Q}_{\overline{c}_i},\mathbb{K}_{\overline{c}_i},\mathbb{V}_{\overline{c}_i})$, $\overline{C}_{local}=\{\overline{c}_{local}^i\}_{i=1}^{T/16}$, $\overline{h}_{local}^j=\text{TRM}_{Local}(\mathbb{Q}_{\overline{h}_j},\mathbb{K}_{\overline{h}_j},\mathbb{V}_{\overline{h}_j})$, $\overline{H}_{local}=\{\overline{h}_{local}^j\}_{j=1}^J$. Thus, the corresponding output of the global transformer is:
\begin{align}
O_{global}^v=\text{TRM}_{Global}(\mathbb{Q}_{\overline{v}_{global}},\mathbb{K}_{\overline{C}_{local}},\mathbb{V}_{\overline{C}_{local}}),\\
O_{global}^q=\text{TRM}_{Global}(\mathbb{Q}_{\overline{q}_{global}},\mathbb{K}_{\overline{H}_{local}},\mathbb{V}_{\overline{H}_{local}}),
\label{out_global}
\end{align}
where $O_{global}^v$ is the visual output of the global transformer and $O_{global}^q$ is the textual output of the local transformer.
Similar to the local transformer, we also add a feed-forward layer and a normalization layer to encode $O_{global}^v$ and $O_{global}^q$. Finally, for each modality, we concatenate the local and global representations to generate the final fine-grained visual/textual features as follows:
\begin{align}
f^v&=concat(O_{local}^v,O_{global}^v),\\
f^q&=concat(O_{local}^q,O_{global}^q),
\label{out_global}
\end{align}
where $f^v$ is the final visual feature and $f^q$ is the final textual feature.

\subsection{Transformer Decoder}
After obtaining the fine-grained visual and textual features $f^v$ and $f^q$, we need a transformer decoder to handle these cross-modal interactions.
Supposing we need to predict $M$ segment candidates, as the additional input, segment queries $\mathbb{S}=\{s_h\}_{h=1}^M$ \cite{wang2022rcl,ding2021decoupling,cao2021pursuit} are utilized to learn a possible segment by aligning the semantics between $f^v$ and $f^q$. Based on $\mathbb{S}$, we develop a cross-modal parallel transformer to  integrate these features from different modalities in parallel. To further assist the multi-modal semantic alignment and interaction, we also design a new cross-modal cycle consistency loss in this decoder for supervision.

\noindent \textbf{Cross-modal parallel transformer.} Given the visual features $f^v$, we employ several linear layers on it to generate a set of video-specific key $\mathbb{K}_{f^v}$ and video-specific value $\mathbb{V}_{f^v}$. Similarly, we can also obtain the query-specific key $\mathbb{K}_{f^q}$ and query-specific value $\mathbb{V}_{f^q}$ as:
\begin{align}
\mathbb{K}_{f^v}&=f^vW_6^{k},\mathbb{V}_{f^v}=f^vW_6^v,\\
\mathbb{K}_{f^q}&=f^qW_7^{k},\mathbb{V}_{f^q}=f^qW_7^v,
    \label{xianxing}
\end{align}
where $W_6^k$, $W_6^v$, $W_7^k$ and $W_7^v$ are learnable parameters.
Based on the modality-specific key and value, we design a modality-specific attention module to fuse multi-modal features by two parallel branches (\textit{i.e.}, two MultiAtt modules) in Fig. \ref{fig:transformer}, where MultiAtt is the standard Multi-head Attention module \cite{vaswani2017attention,xu2022mdan}, which is defined as:
\begin{align}
Att_v&=MultiAtt(\overline{\mathbb{S}},\mathbb{K}_{f^v},\mathbb{Q}_{f^v}),\\
Att_q&=MultiAtt(\overline{\mathbb{S}},\mathbb{K}_{f^q},\mathbb{Q}_{f^q}),
    \label{multiatt}
\end{align}
where $Att_v$ is the attention output in the visual branch and $Att_q$ is the attention output in the textual branch, $\overline{\mathbb{S}}$ denotes the enhanced segment queries by the self-attention operation.  To model fine-grained cross-modal interaction, we integrate these two attentions as follows:
\begin{align}
O_{cross}=Att_v\oplus Att_q,
    \label{atten}
\end{align}
where $\oplus$ denotes the additive sum with learnable weights. $Att_v$ is the sum with learnable weights in the visual branch, and $Att_q$ is the sum with learnable weights in the textual branch.
Note that the main computational cost of the cross-modal parallel transformer is matrix multiplication (\textit{i.e.}, ``matmul'' in Fig. \ref{fig:transformer}).
Based on Eq. \eqref{multiatt} and \eqref{atten}, we can calculate the visual attention and the textual attention in parallel, which improves the computational efficiency.

\noindent \textbf{Cross-modal cycle consistency.} 
In the TSG task, a phrase often corresponds to a specific clip. To enforce better semantic alignment between clips and phrases, we design a new cross-modal cycle-consistency loss during cross-modal interaction. In general, if a clip and a phrase are identified as semantically aligned, their representations are nearest neighbors in the learned common spaces. After obtaining clip-level features $\{\overline{c}_i\}_{i=1}^{T/16}$ or phrase-level features $\{\overline{h}_j\}_{j=1}^J$, we design a cross-modal cycle-consistency constraint for better cross-modal alignment.

Firstly, given phrase $\overline{h}_j$, we find the visual soft nearest neighbor (\textit{i.e.}, the most relevant clip) $\overline{c}_i$ by:
\begin{align}
\overline{c}_i=\sum_{r=1}^{T/16}\frac{\exp{(-||\overline{h}_j-\overline{c}_r||^2)}}{\sum_{t=1}^{T/16}\exp{(-||\overline{h}_j-\overline{c}_t||^2)}}\overline{c}_r,
    \label{ruanlinjv_v}
\end{align}
where ${\exp{(-||\overline{h}_j-\overline{c}_r||^2)}}/{\sum_{t=1}^{T/16}\exp{(-||\overline{h}_j-\overline{c}_t||^2)}}$ is used to compute the similarity score of clip $\overline{c}_i$ and phrase $\overline{h}_j$.

Then, we cycle back from $\overline{c}_i$ to phrase sequence $\{\overline{h}_j\}_{j=1}^J$ by calculating the soft phrase location as follows:
\begin{align}
p=\sum_{j=1}^{J}\frac{\exp{(-||\overline{h}_j-\overline{c}_i||^2)}}{\sum_{\xi=1}^J\exp{(-||\overline{h}_{\xi}-\overline{c}_i||^2)}}j.
    \label{ruanlinjv_q}
\end{align}
To learn semantically consistent representations, we penalize deviations from cycle-consistency for sampled sets of clips and sentences based on the following loss function:
\begin{align}
L_{CMCC}=e^{-\cos(p,j)},
    \label{loss_cmcc}
\end{align}
where $\cos(p,j)$ denotes the cosine similarity of $p$ and $j$.

Finally, by minimizing Eq. \eqref{loss_cmcc}, we can make the source location $j$ and the soft destination location $p$ as close as possible. If $p=j$, the clip $\overline{c}_i$ and the phrase $\overline{c}_i$ are semantically corresponding; if $p \neq j$, we can obtain the nearest neighbor of  $\overline{c}_i$ by minimizing Eq. \eqref{loss_cmcc}.

\subsection{Boundary Prediction}
After obtaining $O_{cross}$ by Eq.~\eqref{atten}, we utilize multiple feed forward networks to obtain a series of fixed-length boundary predictions  $\hat{Y} = \{\hat{y}_h\}_{h=1}^M$, where $\hat{y}_h = (\hat{b}_h; \hat{d}_h)$ contains the $h$-th predicted  segment coordinate   $\hat{b}_{h} \in[0,1]^{2}$ and the corresponding confidence score $\hat{d}_h \in[0, 1]$. We denote
the ground truth as $Y=\{y_h\}_{h=1}^M$, which contains the ground-truth segment coordinate $b \in[0,1]^{2}$.

Based on the fixed-length boundary predictions and ground-truth boundary, we design the set prediction loss as follows:
\begin{align}
L_{boundary}(\hat{y}_{h}, y_h)=\lambda_{\ell_{1}}\|b-\hat{b}_{h}\|_{1}+\lambda_{IoU} \mathcal{C}_{IoU}(b, \hat{b}_{h})-\hat{d}_{h},
    \label{loss_boundary}
\end{align}
where $\lambda_{\ell_{1}}$ and $\lambda_{{IoU}}$ are weighting parameters; $\mathcal{C}_{IoU}(\cdot, \cdot)$ is a scale-invariant generalized intersection over union. By minimizing Eq. \eqref{loss_boundary}, we can determine the optimal prediction segment  slot from multiple boundary predictions. Assuming that the $\mu$-th slot is optimal, we denote the corresponding optimal prediction as $\hat{y}_{\mu}$.

Thus, the final loss is as follows:
\begin{align}
L_{final}=L_{CMCC}+\lambda_fL_{boundary}(\hat{y}_{\mu}. y_h),
    \label{loss_final}
\end{align}
where parameter $\lambda_f$ is utilized to control the balance.

\noindent \textbf{Inference}: The inference process of our proposed HLGT is very simple. Without predefined threshold values or time-consuming post-processing processes, we generate the predicted segment boundary in only one forward pass. The predicted segment  with the highest confidence score will be selected as the final predicted segment.

\section{Experiments}
\subsection{Datasets and Evaluation Metric}
We conduct experiments on three challenging benchmark datasets: ActivityNet Captions \cite{krishna2017dense}, Charades-STA \cite{gao2017tall} and TACoS \cite{regneri2013grounding}.

\noindent \textbf{ActivityNet Captions.}
ActivityNet Captions \cite{krishna2017dense} contains 20k untrimmed videos with 100k language descriptions from YouTube \cite{caba2015activitynet}. These videos are mainly about complicated human activities in daily life.
These videos are 2 minutes on average, and these annotated video segments have much larger variation, ranging from several seconds to over 3 minutes. Since the test split is withheld for competition, following the public split \cite{gao2017tall}, we use 37421, 17505, and 17031 query-video pairs for training, validation, and testing respectively.

\noindent \textbf{Charades-STA.}
Built upon the Charades \cite{sigurdsson2016hollywood} dataset, Charades-STA \cite{gao2017tall} contains 6672 videos and involves 16128 video-query pairs, which pays attention to daily life indoors activities.
It is collected for video action recognition and video captioning, and contains 6672 videos and involves 16128 video-query pairs.
Following \cite{gao2017tall,liu2022exploring}, we utilize 12408 video-query pairs for training and 3720 pairs for testing.

\noindent \textbf{TACoS.}
Collected by \cite{regneri2013grounding} for video grounding and dense video captioning tasks, TACoS  consists of 127 long videos, which are mainly about cooking scenarios. It consists of 127 videos on cooking activities with an average length of 4.79 minutes. In video grounding task, it contains 18818 video-query pairs. For fair comparisons, we follow the same split of the dataset as \cite{gao2017tall}, which has 10146, 4589, and 4083 video-query pairs for training, validation, and testing respectively.

\noindent \textbf{Evaluation metric.}
Following \cite{zhang2020span}, we adopt ``R@$n$, IoU=$m$'' proposed by \cite{hu2016natural} for the metric, which calculates the ratio of at least one of top-$n$ selected segments having an intersection over union (IoU) larger than $m$. The larger metric means the better performance.
In our experiments, we utilize $n \in \{1,5\}$ for all datasets, $m \in \{0.5,0.7\}$ for ActivityNet Captions and Charades-STA, $m \in \{0.3,0.5\}$ for TACoS.

\subsection{Implementation Details}
For video encoding, we define continuous 16 frames as a clip and each clip overlaps 8 frames with adjacent clips. Then, we use a pre-trained Resnet-152 network \cite{he2016deep} to extract the frame-level visual features in each clip.  Since some videos are overlong, we uniformly downsample frame-feature sequences to $T=256$.
For query encoding, we utilize the Glove embedding \cite{PenningtonSM14} to embed each word to 300-dimension features. For both visual and textual branches, we set the hidden state size as 1024 and the attention head as 8 in all the transformer and attention blocks. We train the whole model for 80 epochs with the batch size of 16 and the early stopping strategy. The hyperparameters $\lambda_{\ell_{1}}$, $\lambda_{{IoU}}$ and $\lambda_f$ are set to 0.8, 0.5 and 0.2 respectively according to empirical study.
We perform the parameter optimization  by Adam optimizer \cite{kingma2014adam} with leaning rate $3\times 10^{-4}$ for ActivityNet Captions and Charades-STA and $2\times 10^{-4}$ for TACoS, and linear decay of the learning rate and gradient clipping of 1.0. Our method is implemented by using PyTorch on the machine with two NVIDIA Tesla V100 GPUs.

\subsection{Comparison with State-of-the-Arts}
\noindent \textbf{Compared methods.} We compare the proposed HLTG with state-of-the-art TSG methods on three datasets. These methods are grouped into three categories by the viewpoints of top-down, bottom-up and transformer-based methods.
To make a fair comparison with these TSG methods, following \cite{zhang2021natural}, we cite their results from corresponding works:

(i) Top-down approach: CTRL \cite{gao2017tall}, ACRN \cite{liu2018attentive}, QSPN \cite{xu2019multilevel}, SCDM \cite{yuan2019semantic}, BPNet \cite{xiao2021boundary}, CMIN \cite{zhang2019cross}, 2DTAN \cite{zhang2019learning}, DRN \cite{zeng2020dense}, FIAN \cite{qu2020fine}, CBLN \cite{liu2021context}. These methods first sample multiple candidate video segments, and then directly compute the semantic similarity between the query representations with segment representations for ranking and selection.

(ii) Bottom-up approach: CBP \cite{wang2019temporally}, GDP \cite{chenrethinking}, LGI \cite{mun2020local}, VSLNet \cite{zhang2020span}, IVG-DCL \cite{nan2021interventional}, ACRM \cite{tang2021frame}. These methods directly predict the start and end timestamps of the target segment by regression.

(iii) Transformer-based approach: VIDGTR \cite{woo2022explore}, De-VLTrans-MSA \cite{zhang2021multi}, GTR \cite{cao2021pursuit}. Different from the above top-down and bottom-up approaches, they capture more fine-grained interaction between the video-query input and directly output the segment predictions via the effective transformer encoder-decoder architecture without using any time-consuming pre- and post-processing operation.

\begin{table}[t!]
    \centering
    \caption{Performance compared with the state-of-the-art methods on  ActivityNet Captions, where $\downarrow$ means the top-down setting, $\uparrow$ means the bottom-up setting, and $\leftrightarrow$ means the transformer-based setting.}
    \vspace{-3mm}
    \setlength{\tabcolsep}{1.3mm}{
    \begin{tabular}{c|c|cccc}
    \hline
    \multirow{2}*{Method} & \multirow{2}*{Setting} & R@1, & R@1, & R@5, & R@5, \\
    ~ & ~ & IoU=0.5 & IoU=0.7 & IoU=0.5 & IoU=0.7 \\ \hline \hline
    CTRL \cite{gao2017tall} & $\downarrow$ & 29.01 & 10.34 & 59.17 & 37.54 \\
    ACRN \cite{liu2018attentive}& $\downarrow$ & 31.67 & 11.25 & 60.34 & 38.57 \\
    QSPN \cite{xu2019multilevel}& $\downarrow$ & 33.26 & 13.43 & 62.39 & 40.78 \\
    SCDM \cite{yuan2019semantic}& $\downarrow$ & 36.75 & 19.86 & 64.99 & 41.53 \\
    CMIN \cite{zhang2019cross}& $\downarrow$& 43.40 & 23.88 & 67.95 & 50.73 \\
    2DTAN \cite{zhang2019learning}& $\downarrow$ & 44.51 & 26.54 & 77.13 & 61.96 \\
    DRN \cite{zeng2020dense}& $\downarrow$ & 45.45 & 24.36 & 77.97 & 50.30 \\
    FIAN \cite{qu2020fine}& $\downarrow$ & 47.90 & 29.81 & 77.64 & 59.66 \\
    CBLN \cite{liu2021context}& $\downarrow$ & 48.12 & 27.60 & 79.32 & 63.41 \\ \hline \hline
    CBP \cite{wang2019temporally}& $\uparrow$ & 35.76 & 17.80 & 65.89 & 46.20 \\
    LGI \cite{mun2020local}& $\uparrow$ & 41.51 & 23.07 & - & - \\
    VSLNet \cite{zhang2020span}& $\uparrow$ & 43.22 & 26.16 & - & - \\
    IVG-DCL \cite{nan2021interventional}& $\uparrow$ & 43.84 & 27.10 & - & - \\ \hline \hline
    De-VLTrans-MSA \cite{zhang2021multi}&$\leftrightarrow$&48.02&31.78 &78.02&63.18\\
    GTR \cite{cao2021pursuit}&$\leftrightarrow$& 50.57 & 29.11 & 80.43 & 65.14 \\
    VIDGTR \cite{woo2022explore}&$\leftrightarrow$&53.27& 27.93& 78.19& 57.82\\
    \textbf{Our HLGT} & $\leftrightarrow$ & \textbf{55.68} & \textbf{34.25} & \textbf{84.19} & \textbf{67.43}\\ \hline
    \end{tabular}}
    \label{tab:sota1}
\end{table}

\begin{table}[t!]
    \centering
    \caption{Performance compared with the state-of-the-art methods on  Charades-STA, where $\downarrow$ means the top-down setting, $\uparrow$ means the bottom-up setting, and $\leftrightarrow$ means the transformer-based setting.}
    \vspace{-3mm}
    \setlength{\tabcolsep}{1.3mm}{
    \begin{tabular}{c|c|cccc}
    \hline
    \multirow{2}*{Method} &  \multirow{2}*{Setting} & R@1, & R@1, & R@5, & R@5, \\
    ~ & ~ & IoU=0.5 & IoU=0.7 & IoU=0.5 & IoU=0.7  \\ \hline \hline
    CTRL \cite{gao2017tall}& $\downarrow$ & 23.63 & 8.89 & 58.92 & 29.57  \\
    ACRN \cite{liu2018attentive}& $\downarrow$ & 20.26 & 7.64 & 71.99 & 27.79  \\
     SAP \cite{chen2019semantic} & $\downarrow$ & 27.42 & 13.36 & 66.37 & 38.15  \\
    QSPN \cite{xu2019multilevel}& $\downarrow$ & 35.60 & 15.80 & 79.40 & 45.50  \\
    SCDM \cite{yuan2019semantic}& $\downarrow$ & 54.44 & 33.43 & 74.43 & 58.08 \\
    2DTAN \cite{zhang2019learning}& $\downarrow$ & 39.81 & 23.25 & 79.33 & 51.15  \\
    DRN \cite{zeng2020dense}& $\downarrow$ & 53.09 & 31.75 & 89.06 & 60.05  \\
    FIAN \cite{qu2020fine}& $\downarrow$ & 58.55 & 37.72 & 87.80 & 63.52 \\
    CBLN \cite{liu2021context}& $\downarrow$ & 61.13 & 38.22 & 90.33 & 61.69  \\ \hline \hline
    CBP \cite{wang2019temporally}& $\uparrow$ & 36.80 & 18.87 & 70.94 & 50.19  \\
    GDP \cite{chenrethinking}& $\uparrow$ & 39.47 & 18.49 & - & - \\
    VSLNet \cite{zhang2020span}& $\uparrow$ & 47.31 & 30.19 & - & - \\
    IVG-DCL \cite{nan2021interventional}& $\uparrow$ & 50.24 & 32.88 & - & -  \\
    ACRM \cite{tang2021frame}& $\uparrow$ & 57.53 & 38.33 & - & - \\
    \hline \hline
    VIDGTR \cite{woo2022explore}&$\leftrightarrow$&46.74& 22.72& 84.67& 52.01\\
    GTR \cite{cao2021pursuit}&$\leftrightarrow$& 62.58 & 39.68 & 91.62 & 62.03 \\
    \textbf{Our HLGT} & $\leftrightarrow$ & \textbf{65.31} & \textbf{41.38} & \textbf{94.50} & \textbf{64.72}\\ \hline
    \end{tabular}}
    \label{tab:sota2}
\end{table}

\begin{table}[t!]
    \centering
    \caption{Performance compared with the state-of-the-art methods on TACoS, where $\downarrow$ means the top-down setting, $\uparrow$ means the bottom-up setting, and $\leftrightarrow$ means the transformer-based setting.}
    \vspace{-3mm}
    \setlength{\tabcolsep}{1.3mm}{
    \begin{tabular}{c|c|cccc}
    \hline
    \multirow{2}*{Method} &  \multirow{2}*{Setting} & R@1, & R@1, & R@5, & R@5, \\
    ~ & ~ & IoU=0.3 & IoU=0.5 & IoU=0.3 & IoU=0.5  \\ \hline \hline
    CTRL \cite{gao2017tall}& $\downarrow$ & 18.32 & 13.30 & 36.69 & 25.42 \\
    ACRN \cite{liu2018attentive}& $\downarrow$ & 19.52 & 14.62 & 34.97 & 24.88 \\
    QSPN \cite{xu2019multilevel}& $\downarrow$ & 20.15 & 15.32 & 36.72 & 25.30 \\
    SCDM \cite{yuan2019semantic}& $\downarrow$ & 26.11 & 21.17 & 40.16 & 32.18 \\
    CMIN \cite{zhang2019cross}& $\downarrow$ & 24.64 & 18.05 & 38.46 & 27.02 \\
    2DTAN \cite{zhang2019learning}& $\downarrow$ & 37.29 & 25.32 & 57.81 & 45.03 \\
    FIAN \cite{qu2020fine}& $\downarrow$ & 33.87 & 28.58 & 47.76 & 39.16 \\
    CBLN \cite{liu2021context}& $\downarrow$ & 38.98 & 27.65 & 59.96 & 46.24 \\ \hline \hline
    CBP \cite{wang2019temporally}& $\uparrow$ & 27.31 & 24.79 & 43.64 & 37.40 \\
    VSLNet \cite{zhang2020span}& $\uparrow$ & 29.61 & 24.27 & - & - \\
    IVG-DCL \cite{nan2021interventional}& $\uparrow$ & 38.84 & 29.07 & - & - \\
    ACRM \cite{tang2021frame}& $\uparrow$ & 38.79 & 26.94 & - & - \\
    \hline \hline
    GTR \cite{cao2021pursuit}&$\leftrightarrow$& 40.39 & 30.22 & 61.94 & 47.73\\
    De-VLTrans-MSA \cite{zhang2021multi}&$\leftrightarrow$& 48.79 & 37.57 & 67.63 & 57.91\\
    \textbf{Our HLGT} & $\leftrightarrow$ & \textbf{49.33} & \textbf{39.17} & \textbf{69.06} & \textbf{58.94} \\ \hline
    \end{tabular}}
    \label{tab:sota3}
\end{table}

\noindent \textbf{Comparison on ActivityNey Captions.}
We compare our proposed HLGT with the state-of-the-art top-down, bottom-up, and transformer-based TSG methods on the ActivityNey Captions dataset in Table~\ref{tab:sota1}, where our HLGT reaches the best performance over all the metrics. Particularly, compared with the best top-down approach CBLN, our HLGT achieves 7.56\%, 6.65\%, 4.87\% and 4.02\% improvements on  all the metrics, respectively. Our  HLGT also obtains an even larger improvement over the best bottom-up method IVG-DCL in metrics R@1, IoU=0.5 and R@1, IoU=0.5 by 11.84\% and 7.15\%. It verifies the benefits of utilizing our end-to-end architecture to  effectively model the fine-
grained visual-language alignment between video and language query.
Besides, HLGT outperforms other transform-based methods by a large margin in all the metrics. Particularly, compared to the best transformer-based method GTR, our proposed HLGT brings the improvement by 5.51\%, 5.14\%, 3.76\% and 2.29\% in all the metrics.
The significant improvement demonstrates the effectiveness of our multi-level interaction network and brand-new cross-modal cycle-consistency loss.

\noindent \textbf{Comparison on Charades-STA.}
As shown in Table \ref{tab:sota2}
We also compare our proposed HLGT with the state-of-the-art top-down, bottom-up, and transformer-based TSG methods on the Charades-STA dataset. Obviously, our HLGT beats all the other methods  over all evaluation metrics. Compared to the best top-down method CBLN, our HLGT outperforms it by 4.18\% and 4.17\% absolute improvement in terms of R@1, IoU=0.5 and R@5, IoU=0.5, respectively. Compared to the best bottom-up method ACRM, our HLGT improves the performance by 7.78\% and 3.05\% in terms of R@1, IoU=0.5 and R@1, IoU=0.7, respectively. Besides, our HLGT  beats the best transformer-based method GTR by 2.73\% and 2.88\%  absolute improvement in terms of R@1, IoU=0.5 and R@5, IoU=0.5, respectively.

\noindent \textbf{Comparison on TACoS.}
To further compare our proposed HLGT with the state-of-the-art top-down, bottom-up, and transformer-based TSG methods, we  present the results in Table \ref{tab:sota3}. We can find that HLGT still outperforms all the other TSG methods in terms of all the metrics.
Compared to the best top-down method CBLN, our  HLGT outperforms it by 10.35\%, 11.52\%, 9.1\% and 12.7\% in terms of all metrics, respectively. HLGT also beats the best bottom-up method ACRM and brings the improvements by 10.54\% and 12.13\% in terms of R@1, IoU=0.3 and R@1, IoU=0.5, respectively. Compared to the best transformer-based method De-VLTrans-MSA,  HLGT brings the improvements of 1.60\% and 1.43\% in terms of R@1, IoU=0.5 and R@5, IoU=0.3, respectively.

\begin{figure}[t!]
\centering
\includegraphics[width=0.223\textwidth]{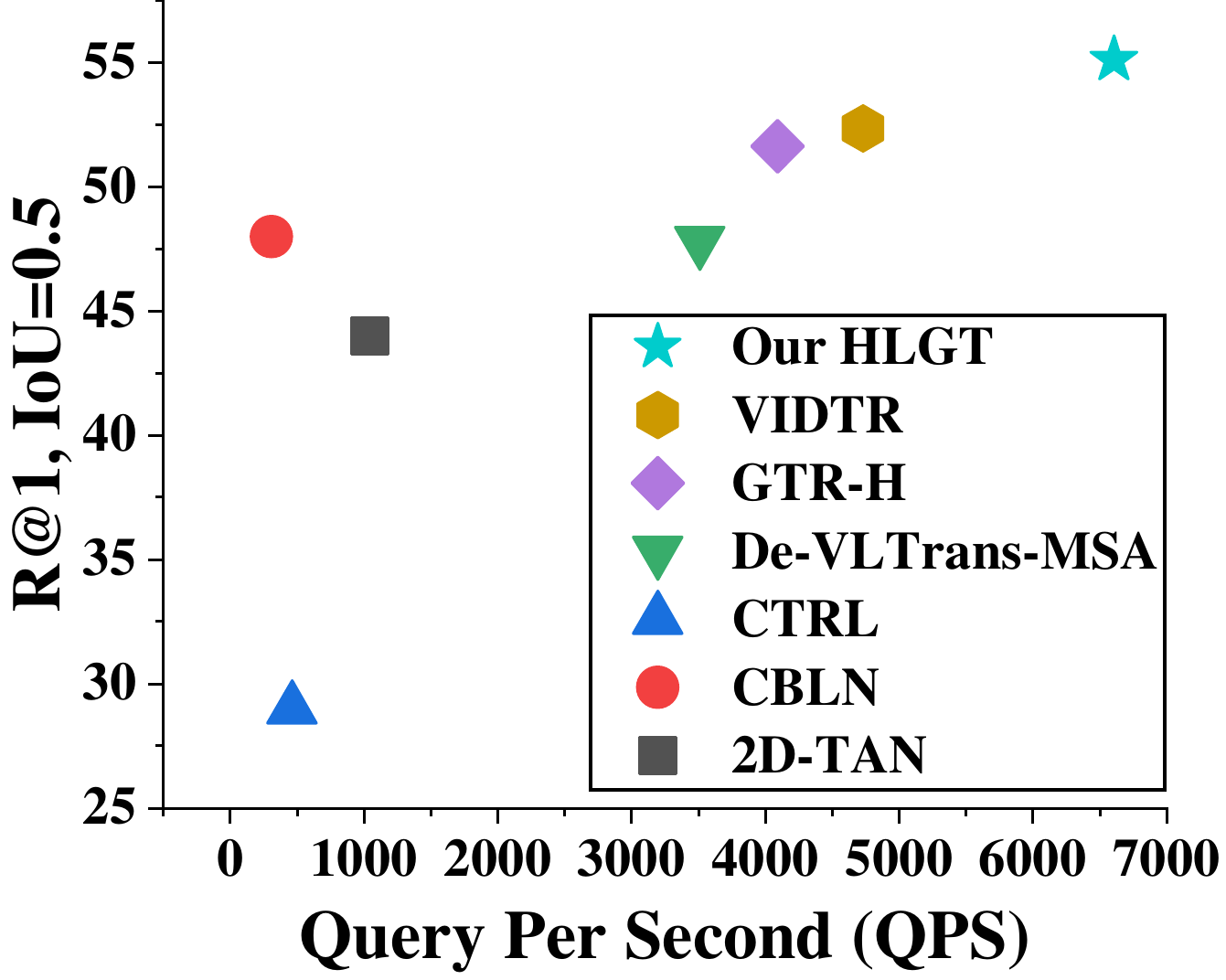}
\includegraphics[width=0.223\textwidth]{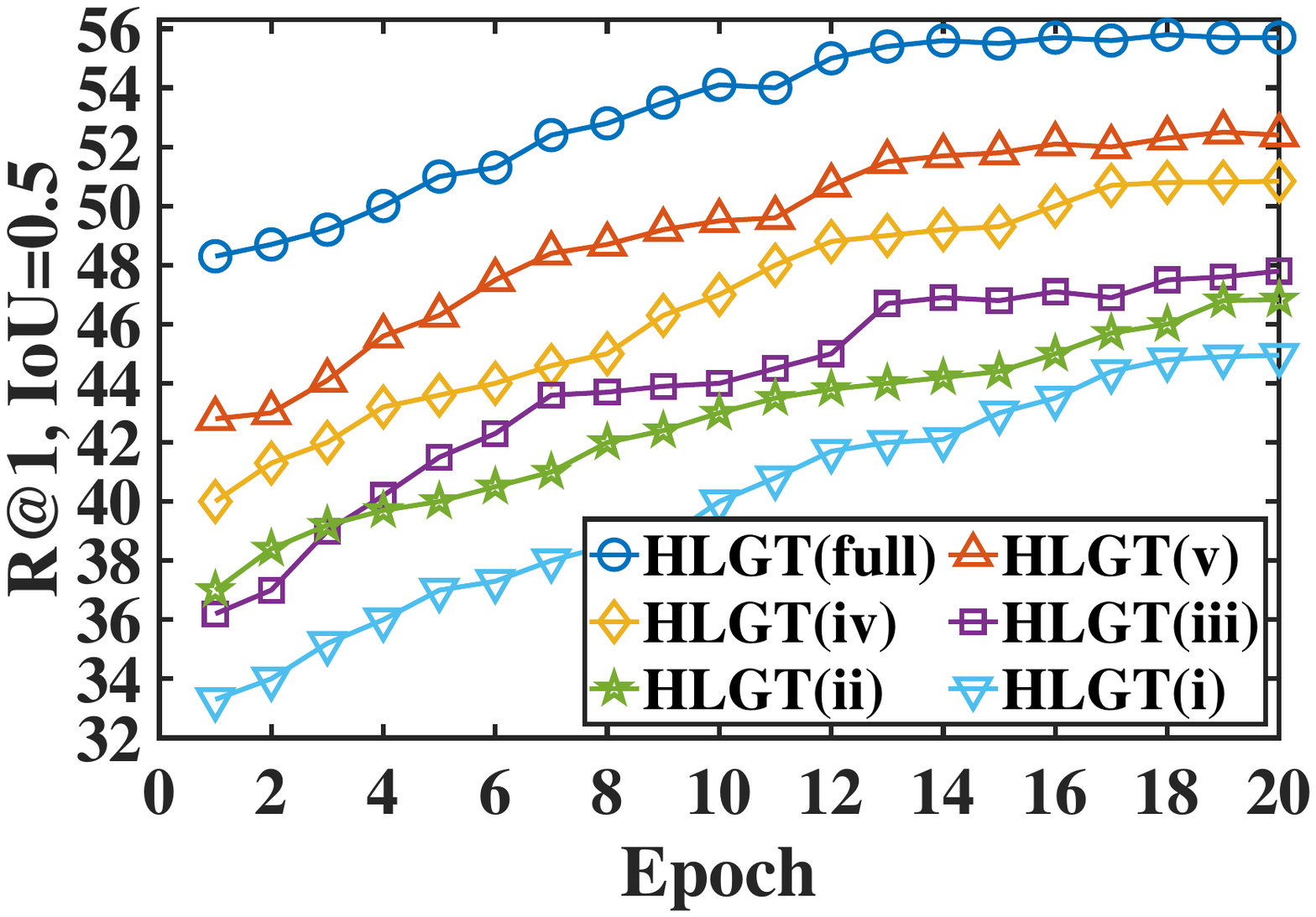}
\caption{Performance comparisons on  ActivityNet Captions. Left: Efficiency comparisons in terms of R@1, IoU=0.5 and  query per second (QPS, \textit{i.e.}, the number of localized queries per second). Right: Performance of each ablation module during training epochs.}
\vspace{-3mm}
\label{fig:efficiency}
\end{figure}

\noindent \textbf{Efficiency comparison.}
We further evaluate the efficiency of our proposed HLGT by fairly comparing its inference speed (QPS) with state-of-the-art methods on the challenging ActivityNet Captions dataset in Fig.~\ref{fig:efficiency}. HLGT can process more than $6\times 10^3$  queries per second, which shows that our HLGT can efficiently process these challenging multi-modal data. Compared with other state-of-the-art methods, our HLGT runs faster and achieves better grounding performance. Particularly, our HLGT is 5.24\% better than transformer-based method VIDTR with 38.71\% faster speed, which shows the effectiveness and efficiency of our HLGT. Our satisfactory performance attributes to: (i) Our cross-modal parallel transformer  is able to process visual features and textual features in parallel, which effectively reduces the time consumption of processing multi-modal features. (ii) For each modality, our global-local transformer can learn the interactions between the local and global semantics for better multi-modal reasoning and more accurate video grounding. Therefore, our HLGT will have wider real-world multimedia applications, due to its efficiency and effectiveness on the challenging large-scale  ActivityNet Captions dataset.

\subsection{Ablation Study}
To examine the effectiveness of each component in our HLTG, we perform in-depth ablation studies on three challenging datasets: ActivityNet Captions, Charades-STA and TACoS.

\begin{table*}[th]
    \centering
    \caption{Main ablation studies of our proposed HLGT on all three datasets, which investigates feature extraction, transformer encoder, transformer decoder, and boundary prediction.}
    \vspace{-3mm}
    \setlength{\tabcolsep}{1.3mm}{
    \begin{tabular}{c|cc|cc|c|c|ccccccccccccc}
    \hline
    \multicolumn{11}{c}{ActivityNet Captions}\\\hline
    \multirow{2}*{Model} & \multicolumn{2}{c}{Feature Extraction} &  \multicolumn{2}{|c|}{Transformer Encoder}& Transformer Decoder &  Boundary Prediction  & R@1, & R@1, &R@5, & R@5, \\ \cline{2-7}
    ~ &  Local & Global & Local & Global&$L_{CMCC}$ &$L_{boundary}$ &IoU=0.5 & IoU=0.7 &IoU=0.5 & IoU=0.7 \\ \hline
    i & $\times$ & $\times$ & $\times$ &  $\times$ & $\times$ & $\times$  & 44.95&28.37&75.08&60.72 \\
    ii & $\checkmark$ & $\times$ & $\checkmark$ &  $\times$ &  $\times$ &$\times$  &46.83&30.21&78.92&63.14 \\
    iii & $\times$ & $\checkmark$ & $\times$ &  $\checkmark$  & $\times$ & $\times$ & 47.56&29.33&79.31&62.05\\
   iv & $\checkmark$ & $\checkmark$ & $\checkmark$ &  $\checkmark$ &  $\times$ & $\times$ & 50.84&30.12 &81.34&63.18\\
    v & $\checkmark$ & $\checkmark$ & $\checkmark$ &  $\checkmark$ &  $\checkmark$ & $\times$ &52.41&31.07&82.56&64.32  \\
    \textbf{Full}  & $\checkmark$ & $\checkmark$ & $\checkmark$ &  $\checkmark$ & $\checkmark$ & $\checkmark$ &\textbf{55.68} & \textbf{34.25} & \textbf{84.19} & \textbf{67.43} \\ \hline \hline
    \multicolumn{11}{c}{TACoS}\\\hline
    \multirow{2}*{Model} & \multicolumn{2}{c}{Feature Extraction} &  \multicolumn{2}{|c|}{Transformer Encoder}& Transformer Decoder &  Boundary Prediction  & R@1, & R@1, &R@5, & R@5, \\ \cline{2-7}
    ~ &  Local & Global & Local & Global&$L_{CMCC}$ &$L_{boundary}$  &IoU=0.3 & IoU=0.5 &IoU=0.3 & IoU=0.5 \\ \hline
    i & $\times$ & $\times$ & $\times$ &  $\times$ & $\times$ & $\times$  & 42.95&32.76&60.32&50.20 \\
    ii & $\checkmark$ & $\times$ & $\checkmark$ &  $\times$ &  $\times$ &$\times$  & 44.36&35.71&62.18&51.42\\
    iii & $\times$ & $\checkmark$ & $\times$ &  $\checkmark$  & $\times$ & $\times$ &45.17&36.02&63.71&52.04 \\
   iv & $\checkmark$ & $\checkmark$ & $\checkmark$ &  $\checkmark$ &  $\times$ & $\times$ &46.31&36.95&64.88&53.76  \\
    v & $\checkmark$ & $\checkmark$ & $\checkmark$ &  $\checkmark$ &  $\checkmark$ & $\times$ & 47.64&37.83&66.98&56.10 \\
    \textbf{Full}  & $\checkmark$ & $\checkmark$ & $\checkmark$ &  $\checkmark$ & $\checkmark$ & $\checkmark$ &\textbf{49.33} & \textbf{39.17} & \textbf{69.06} & \textbf{58.94} \\  \hline \hline
    \multicolumn{11}{c}{Charades-STA}\\\hline
    \multirow{2}*{Model} & \multicolumn{2}{c}{Feature Extraction} &  \multicolumn{2}{|c|}{Transformer Encoder}& Transformer Decoder &  Boundary Prediction  & R@1, & R@1, &R@5, & R@5, \\ \cline{2-7}
    ~ &  Local & Global & Local & Global&$L_{CMCC}$ &$L_{boundary}$  &IoU=0.5 & IoU=0.7 &IoU=0.5 & IoU=0.7  \\ \hline
    i & $\times$ & $\times$ & $\times$ &  $\times$ & $\times$ & $\times$  &58.45&34.67&87.19&59.94  \\
    ii & $\checkmark$ & $\times$ & $\checkmark$ &  $\times$ &  $\times$ &$\times$  &60.24&37.05&90.06&61.13 \\
    iii & $\times$ & $\checkmark$ & $\times$ &  $\checkmark$  & $\times$ & $\times$ &59.43&36.72&89.44&60.87 \\
   iv & $\checkmark$ & $\checkmark$ & $\checkmark$ &  $\checkmark$ &  $\times$ & $\times$ &61.87&38.54&90.71&62.35  \\
    v & $\checkmark$ & $\checkmark$ & $\checkmark$ &  $\checkmark$ &  $\checkmark$ & $\times$ & 64.18&40.02&92.41&63.26 \\
    \textbf{Full}  & $\checkmark$ & $\checkmark$ & $\checkmark$ &  $\checkmark$ & $\checkmark$ & $\checkmark$ &\textbf{65.31} & \textbf{41.38} & \textbf{94.50} & \textbf{64.72} \\ \hline
    \end{tabular}}
    \label{tab:main_ablation_three}
\end{table*}

\noindent \textbf{Main ablation study.}
We first conduct the main ablation study to examine the effectiveness of all the modules in our model, including multi-level feature extractions (local and global), multi-level transformer encoders (local and global),  the transformer decoder  and the boundary predict module.
The ablation results are reported in Table~\ref{tab:main_ablation_three}:
1) Model i is the baseline model without temporal transformer and feature fusion, where we directly employ these frame- and word-level features for grounding.
2) For each modality, Model ii only uses the local features and ignores the global features for grounding. 3) On the contrary, Model iii only utilizes the global features for grounding. 4) As for Model iv, we use both local and global features for grounding. 5) In Model v, we add the CMCC loss to Model iii. 6) Model Full is our full HLGT.

From Table~\ref{tab:main_ablation_three}, we can find that: (i) Model Full performs the best and Model i the worst.
(ii) Compared to Model i, Model ii and iii achieve the improvement by  1.18\% and 2.61\% respectively in terms of ``R@1, Iou=0.5'' on the ActivityNet Captions dataset. It shows that local  and global features can be used to align visual and textual representations.
(iii) Compared to Model ii and iii, Model iv improves the performance by 1.49\% and 1.82\% respectively in terms of ``R@1, Iou=0.7'' on the Charades-STA dataset. It is because both local  and global features are  significant to learn the full visual/textual representation. (iv) As for Model v, it outperforms Model iv by 2.23\% in terms of ``R@5, Iou=0.5'' on the TACoS dataset. It is because our cross-modal cycle consistency can encourage semantic alignment between visual and language features in the joint embedding space for video grounding.
(v) Compared to Model v, Model Full achieves the performance improvement by 1.36\% in terms of ``R@1, Iou=0.7'' on the Charades-STA dataset.

\noindent \textbf{Training process of different ablation models.} Following \cite{lin2020weakly}, we try to analyze the training process and grounding performance of different ablation models. Fig. \ref{fig:efficiency} shows the experimental results. We can obtain the following representative observations:
(i) On each epoch, HLGT(full) outperforms other ablation models, which demonstrates the effectiveness of each module. For example, compared to the second-best model HLGT(v), HLGT(full) improves the performance by 3.27\%.
(ii) HLGT(full) converges faster than ablation models, which shows that our full model is more efficient on time-consuming. For instance, HLGT(full) converges within 14 epochs, while HLGT(i) converges after 18 epochs. Thus, our full HLGT can process these challenging datasets more efficiently.

\begin{table}[t!]
   \centering
    \caption{Ablation study on visual feature extractor network on  ActivityNet Captions.}
    \vspace{-3mm}
    \setlength{\tabcolsep}{1.3mm}{
    \begin{tabular}{c|c|c|c|cc}
\hline
\multirow{2}*{Option}  & R@1,   & R@1,   & R@5, & R@5, \\
&IoU=0.5&IoU=0.7&IoU=0.5&IoU=0.7\\\hline
VIDGTR(C3D) & 53.27& 27.93& 78.19& 57.82\\
GTR(C3D)&49.04 & 28.65& 79.26& 64.04 \\
\textbf{Our HLGT(C3D)} &\textbf{55.19}&\textbf{33.84}&\textbf{83.77}&\textbf{67.13}  \\\hline
VIDGTR(I3D)  & 53.55& 28.07& 78.52&58.13\\
GTR(I3D)& 49.32& 28.96& 80.01&64.83 \\
\textbf{Our HLGT(I3D)} & \textbf{55.25}&\textbf{33.90}&\textbf{83.92}&\textbf{67.24}\\\hline
VIDGTR(Plain Transformer)  & 50.39& 28.32&79.14&58.39 \\
GTR(Plain Transformer)& 50.57& 29.11& 80.43& 65.14\\
\textbf{Our HLGT(Plain Transformer)} &\textbf{55.01}&\textbf{33.72}&\textbf{83.59}&\textbf{67.02} \\\hline
VIDGTR(Resnet-152) & 49.69& 27.94&78.36&58.70\\
GTR(Resnet-152)& 50.13&28.76&80.02&64.51 \\
\textbf{Our HLGT(Resnet-152)}  & \textbf{55.68} & \textbf{34.25} & \textbf{84.19} & \textbf{67.43}\\\hline
   \end{tabular}}
    \label{tab:resnet}
\end{table}

\noindent \textbf{Analysis on different visual feature extractor network.} Most of previous methods use pre-trained C3D or I3D to obtain the visual features. However, both C3D and I3D only obtain the clip-level features not the frame-level features. Different from them, we utilize a Resnet-152 network to obtain the fine-grained frame-level feature. In this subsection, we conduct an ablation study experiment to analyze the performance of our used Resnet-152 network. Table~\ref{tab:resnet} shows the results, where the plain transformer is the baseline network. Obviously, our proposed HLGT on the Resnet-152 network performs better than other clip-level pre-trained feature extractor network (C3D and I3D).
Specifically, compared to HLGT(C3D), our used HLGT(Resnet-152) significantly improves the grounding performance by 0.39\%, 0.41\%, 0.42\% and 0.30\% over all metrics. The satisfactory performance improvement illustrates the effectiveness of our used Resnet-152 network.

\begin{table}[t!]
   \centering
    \caption{Performance comparison of different transformers  in terms of R@1, video per second (VPS) and parameters (Para.), where \ding{172} is our  temporal transformer, \ding{173} is our   global-local transformer, and \ding{174} is our  cross-modal parallel transformer. In our experiments, we utilize a transformer module from ``Option'' to replace a transformer module (\ding{172}, \ding{173} and \ding{174}) in our HLGT with freezing    our remaining two transformer modules.}
    \vspace{-3mm}
    \setlength{\tabcolsep}{1.3mm}{
    \begin{tabular}{c|c|cc|cccccc}
    \toprule
\multicolumn{6}{c}{Different Transformer Modules on  TACoS}  \\ \midrule
   {Replaced} & \multirow{2}*{Option} & R@1, & R@1,& \multirow{2}*{VPS $\uparrow$} &\multirow{2}*{Para. $\downarrow$}\\
    Module & ~ & IoU=0.3 & IoU=0.5 \\ \midrule
    \multirow{3}*{\ding{172}} & TokShif \cite{zhang2021token}&46.37&35.54&93.54&291\\
    ~ & MVDeTr \cite{hou2021multiview}&48.02&38.21&126.75&295 \\
~&ViTAE \cite{zhang2022vitaev2}&49.03&38.76&104.79&286\\
 \midrule
    \multirow{3}*{\ding{173}} & Local Transformer \cite{vaswani2017attention}& 42.51& 30.87&82.51& \textbf{83} \\
 ~ & Global Transformer \cite{dosovitskiy2020image}&44.87&31.79&117.28&251\\
    ~&SGLANet \cite{min2020isia}&48.43&35.76&132.35&214\\\midrule
    \multirow{3}*{\ding{174}} & GTR \cite{cao2021pursuit} &41.08& 31.28&153.75&179  \\
  ~& VIDGTR \cite{woo2022explore}&42.86&33.15&164.85&139\\
  ~&De-VLTrans-MSA \cite{zhang2021multi}&48.92&38.05&176.48&143\\\midrule
    - & \textbf{Ours} & \textbf{49.33}& \textbf{39.17} & \textbf{196.54}& 116 \\\bottomrule
   \end{tabular}
\begin{tabular}{c|c|c|c|cc}
\hline\hline
\multicolumn{6}{c}{Different Transformer Layers on  Charades-STA}  \\\hline
\multirow{2}*{Component}&\multirow{2}*{Option}  & R@1,   & R@1,   &\multirow{2}*{VPS $\uparrow$} &\multirow{2}*{Para. $\downarrow$} \\
&&IoU=0.5&IoU=0.7\\\hline
Temporal &Layer=1 &65.31 & 41.38&\textbf{257.38}&\textbf{114} \\
Transformer&Layer=2 &65.37&41.42&210.36&143 \\
&Layer=3  &\textbf{65.42}&\textbf{41.46}&192.47&169 \\\hline
Global-local &Layer=1 &65.31 & 41.38&\textbf{257.38}&\textbf{114} \\
Transformer&Layer=2  & 65.42&41.86&203.59&158\\
&Layer=3  & \textbf{65.47}&\textbf{41.93}&179.87&184\\\hline
Cross-modal Parallel &Layer=1 &65.31 & 41.38&\textbf{257.38}&\textbf{114} \\
Transformer&Layer=2  &65.35&41.39&214.85&138 \\
&Layer=3  &\textbf{65.39}&\textbf{41.45}&185.46&173 \\\hline
   \end{tabular}
   }
    \label{tab:ablation_transformer}
\end{table}

\noindent \textbf{Effect of different transformers.}
To examine the effect of our transformers (\textit{i.e.}, three designed transformers in Fig. \ref{fig:transformer}), we replace our designed transformers with some state-of-the-art transformer modules. For this ablation study, we consider two aspects: different transformer modules on TACoS and different transformer layers on Charades-STA\footnote{Since TACoS is the most challenging dataset, we choose it to better compare the performance of different transformer modules. Considering that Charades-STA is a large-scale dataset, we utilize it to show the performance difference for different transformer layers.}. Table~\ref{tab:ablation_transformer} shows the experimental results.

For different transformer modules on  TACoS, compared to these state-of-the-art transformer modules, our transformers perform better (R@1) and run faster (VPS) in all cases. Particularly, for module \ding{172}, our temporal transformer uses far fewer parameters and runs faster (larger VPS) than other three transformers (TokShif, MVDeTr, and ViTAE). Besides, our temporal transformer significantly outperforms them, where our temporal transformer beats the second-best transformer ViTAE by 0.41\% in terms of ``R@1, IoU=0.5''. As for module \ding{173}, compared to other transformers, our global-local transformer obtains significant performance improvement by a large margin. Especially, our global-local transformer gets better performance than the second-best transformer SGLANet by 3.41\% in terms of ``R@1, IoU=0.5''. For module \ding{174}, our cross-modal parallel transformer beats other transformers over all the metrics. For instance, compared to the second-best transformer De-VLTrans-MSA, our cross-modal parallel transformer achieves performance improvement by 1.12\% in terms of ``R@1, IoU=0.5''. The satisfactory performance of our transformers shows the effectiveness of our transformers, each of which contributes to the model  performance.

About the impact of different  transformer layers, it shows that for each component, the multi-layer transformer only performs marginally better than the single-layer transformer with the higher computational cost (smaller VPS and larger Para.) of transformer operations. An interesting finding is that the multi-layer global-local transformer can bring more improvement than the other two transformers. It is because the global-local transformer can effectively integrate multi-grained features, which improves the grounding performance. For the temporal transformer and the cross-modal parallel transformer, ``layer=1'' achieves  similar performance compared to ``layer=3'' but significantly decreases the computation (Para.). Therefore, for all the transformers, we set  ``layer=1'', which is the suggested value on our TSG task.

\begin{table}[t!]
   \centering
    \caption{Ablation study on segment queries on  Charades-STA.}
    \vspace{-3mm}
    \setlength{\tabcolsep}{1.3mm}{
    \begin{tabular}{c|c|c|c|cc}
\hline
\multirow{2}*{Option}  & R@1,   & R@1,   & R@5, & R@5, \\
&IoU=0.5&IoU=0.7&IoU=0.5&IoU=0.7\\\hline
wo./ Segment Queries & 64.88&41.02&94.37&64.48\\
\textbf{w./ Segment Queries} & \textbf{65.31} & \textbf{41.38} & \textbf{94.50} & \textbf{64.72}\\\hline
   \end{tabular}}
    \label{tab:segment_queries}
\end{table}
\begin{table}[t!]
   \centering
    \caption{Ablation study on shared weight on Charades-STA.}
    \vspace{-3mm}
    \setlength{\tabcolsep}{1.3mm}{
    \begin{tabular}{c|c|c|c|cc}
\hline
\multirow{2}*{Component}&\multirow{2}*{Option}  & R@1,   & R@1,   & R@5, & R@5, \\
&&IoU=0.5&IoU=0.7&IoU=0.5&IoU=0.7\\\hline
Temporal &Unshared Weight & 64.94&41.02&94.18&64.57\\
Transformer&\textbf{Our Shared Weight}  & \textbf{65.31} & \textbf{41.38} & \textbf{94.50} & \textbf{64.72} \\\hline
Feature &Unshared Weight &64.82&40.99&94.02& 64.26\\
Fusion&\textbf{Our Shared Weight}  & \textbf{65.31} & \textbf{41.38} & \textbf{94.50} & \textbf{64.72}\\\hline
   \end{tabular}}
    \label{tab:shared_weight}
\end{table}
\begin{table}[t!]
   \centering
    \caption{Choices for cross-modal cycle consistency on all three datasets.}
    \vspace{-3mm}
        \setlength{\tabcolsep}{1.3mm}
    \resizebox{\linewidth}{!}{
    \begin{tabular}{@{}c|cccccc@{}}
        \toprule
        \multicolumn{5}{c}{Charades-STA}\\\midrule
        \textbf{Loss} & R@1, IoU=0.5 & R@1, IoU=0.7 & R@5, IoU=0.5 & R@5, IoU=0.7\\
        \midrule
        $L_1$  &60.14&37.57&90.42&61.31\\
        $L_2$  & 63.95&38.14&91.18&62.70\\
        $L_3$  & 64.18&40.71&93.56&63.21\\
        \textbf{Our $L_{CMCC}$} &  \textbf{65.31} & \textbf{41.38} & \textbf{94.50} & \textbf{64.72} \\
        \bottomrule \bottomrule
        \multicolumn{5}{c}{ActivityNet Captions}\\\midrule
        \textbf{Loss} & R@1, IoU=0.5 & R@1, IoU=0.7 & R@5, IoU=0.5 & R@5, IoU=0.7\\
        \midrule
        $L_1$  &54.92&33.86&82.74&66.83\\
        $L_2$  &54.86&33.97&83.05&67.21 \\
        $L_3$  &55.09&34.18&83.92&66.84 \\
        \textbf{Our $L_{CMCC}$} & \textbf{55.68} & \textbf{34.25} & \textbf{84.19} & \textbf{67.43}  \\
        \bottomrule \bottomrule
        \multicolumn{5}{c}{TACoS}\\\midrule
        \textbf{Loss} & R@1, IoU=0.3 & R@1, IoU=0.5 & R@5, IoU=0.3 & R@5, IoU=0.5\\
        \midrule
        $L_1$  &49.06&38.75&68.34&57.91\\
        $L_2$  &48.95&39.02&68.52&58.03 \\
        $L_3$  & 49.17&39.04&68.34&58.26\\
        \textbf{Our $L_{CMCC}$} & \textbf{49.33} & \textbf{39.17} & \textbf{69.06} & \textbf{58.94}  \\
        \bottomrule
    \end{tabular}
    }
    \label{tab:cmcc}
\end{table}

\noindent \textbf{Impact of segment queries.} To investigate the effectiveness of segment queries on the transformer decoder, we conduct the ablation study on the Charades-STA dataset. We can find that with segment queries, our HLGT achieves the significant performance improvement. Specifically, compared with the first option (wo./ segment queries), the second option (w/ segment queries) improves the performance by 0.43\%, 0.36\%, 0.13\% and 0.24\% over all metrics, respectively. The significant performance improvement shows the effectiveness of our used segment queries.

\noindent \textbf{Impact of shared weight.} To analyze the impact of shared weight in temporal transformer and feature fusion components, we conduct an ablation study on the Charades-STA dataset. For the setting of unshared weight, we assign an independent weight to each temporal transformer and feature fusion component. As shown in Table \ref{tab:shared_weight}, by introducing the shared weight approach, our HLGT can significantly improve the grounding performance. Compared to the setting of unshared weight, for the temporal transformer, our setting of shared weight achieves the performance improvement by 0.37\% in terms of ``R@1, IoU=0.5''. As for the feature fusion, our setting improves the performance by 0.49 \% in terms of ``R@1, IoU=0.5''. The improvement shows the effectiveness of the shared weight in temporal transformer and feature fusion components.

\begin{table*}[t!]
   \centering
    \caption{Effect of different hyperparameters $\lambda_{\ell_{1}}$, $\lambda_{{IoU}}$ and $\lambda_f$ on all three datasets.}
    \vspace{-3mm}
    \begin{tabular}{c|cccc|cccc|ccccccc}
        \toprule
        \multirow{3}*{$\lambda_{\ell_{1}}$}&\multicolumn{4}{c}{ActivityNet Captions}&\multicolumn{4}{|c|}{Charades-STA}&\multicolumn{4}{c}{TACoS}\\\cline{2-13}
        ~& R@1, & R@1,&R@5,&R@5, & R@1, & R@1,&R@5,&R@5,& R@1, & R@1,&R@5,&R@5,\\
        ~&IoU=0.5 & IoU=0.7 &  IoU=0.5 &  IoU=0.7&  IoU=0.5 &  IoU=0.7 &  IoU=0.5 &  IoU=0.7&  IoU=0.3 &  IoU=0.5 &  IoU=0.3 &  IoU=0.5\\
        \midrule
        0.4  & 53.82&31.69&82.94&65.07&61.33&39.17&92.49&61.93&46.52&37.28&66.49&55.70\\
        0.6  & 55.01&32.98&83.72&66.15&62.95&40.86&93.01&62.99&48.27&38.04&68.15&57.13\\
        \textbf{0.8} &  \textbf{55.68} & \textbf{34.25} & \textbf{84.19} & \textbf{67.43}& \textbf{65.31} & \textbf{41.38} & \textbf{94.50} & \textbf{64.72}& \textbf{49.33} & \textbf{39.17} & \textbf{69.06} & \textbf{58.94}  \\
        1&54.13&32.76&83.04&65.91&63.49&41.25&92.48&63.10&49.12&38.76&67.65&57.09\\\bottomrule\bottomrule
        \multirow{3}*{$\lambda_{{IoU}}$}&\multicolumn{4}{c}{ActivityNet Captions}&\multicolumn{4}{|c|}{Charades-STA}&\multicolumn{4}{c}{TACoS}\\\cline{2-13}
        ~& R@1, & R@1,&R@5,&R@5, & R@1, & R@1,&R@5,&R@5,& R@1, & R@1,&R@5,&R@5,\\
        ~&IoU=0.5 & IoU=0.7 &  IoU=0.5 &  IoU=0.7&  IoU=0.5 &  IoU=0.7 &  IoU=0.5 &  IoU=0.7&  IoU=0.3 &  IoU=0.5 &  IoU=0.3 &  IoU=0.5\\
        \midrule
        0.3&63.25&31.68&81.05&64.30&61.89&38.72&91.37&61.04&46.21&36.28&66.09&55.32\\
        0.4&54.16&33.72&82.64&65.75&63.82&39.67&93.51&63.28&48.39&37.50&67.82&56.73\\
        \textbf{0.5}& \textbf{55.68} & \textbf{34.25} & \textbf{84.19} & \textbf{67.43}& \textbf{65.31} & \textbf{41.38} & \textbf{94.50} & \textbf{64.72}& \textbf{49.33} & \textbf{39.17} & \textbf{69.06} & \textbf{58.94}\\
        0.6&55.14&32.78&82.95&66.03&63.12&40.84&92.87&62.34&47.52&38.25&67.31&57.93\\\bottomrule\bottomrule
        \multirow{3}*{$\lambda_f$}&\multicolumn{4}{c}{ActivityNet Captions}&\multicolumn{4}{|c|}{Charades-STA}&\multicolumn{4}{c}{TACoS}\\\cline{2-13}
        ~& R@1, & R@1,&R@5,&R@5, & R@1, & R@1,&R@5,&R@5,& R@1, & R@1,&R@5,&R@5,\\
        ~&IoU=0.5 & IoU=0.7 &  IoU=0.5 &  IoU=0.7&  IoU=0.5 &  IoU=0.7 &  IoU=0.5 &  IoU=0.7&  IoU=0.3 &  IoU=0.5 &  IoU=0.3 &  IoU=0.5\\
        \midrule
        0.1&54.07&33.85&83.69&66.02&64.17&40.02&92.11&63.02&48.15&38.06&67.92&56.32\\
        \textbf{0.2}&  \textbf{55.68} & \textbf{34.25} & \textbf{84.19} & \textbf{67.43}& \textbf{65.31} & \textbf{41.38} & \textbf{94.50} & \textbf{64.72}& \textbf{49.33} & \textbf{39.17} & \textbf{69.06} & \textbf{58.94}\\
        0.3&54.87&33.72&83.05&66.31&64.07&40.18&93.04&63.18&47.30&37.95&68.01&57.37\\
        0.4&52.10&31.74&82.06&64.18&62.50&38.42&91.03&61.84&46.35&36.72&66.51&55.40\\
        \bottomrule
    \end{tabular}
    \label{tab:chaocan}
\end{table*}
\begin{table}[t!]
   \centering
    \caption{Video per second (VPS) and parameters (Para.) on all three datasets.}
    \vspace{-3mm}
    \setlength{\tabcolsep}{1.3mm}{
    \begin{tabular}{c|c|c|c}
\hline
Dataset  & Domain&VPS & Para.\\\hline
ActivityNet Captions&Open & 190.76 & 119\\\hline
Charades-STA &Indoor& 257.38 & 114\\\hline
TACoS &Cooking& 196.54 & 116\\\hline
   \end{tabular}}
    \label{tab:three_data_performance}
\end{table}

\noindent \textbf{Choices for CMCC loss.} Cross-modal cycle consistency (CMCC) is a significant assistance to enforce the cross-modal interaction. We compare our proposed CMCC loss ($L_{CMCC}$ in Eq. \eqref{loss_cmcc}) with three following popular loss functions: 1) L1 loss $L_1=||p-j||_1^2$; 2) L2 loss $L_2=||p-j||_2^2$; 3) cosine similarity loss $L_3=-cos(p,j)$. Table~\ref{tab:cmcc} reports the results on all the datasets. Obviously, $L_{CMCC}$ dramatically improves grounding performance than other losses, which illustrates the effectiveness of our $L_{CMCC}$.
Especially, compared to the second-best loss $L_3$, our proposed CMCC loss improves the performance by 1.68\% in terms of ``R@5, IoU=0.5'' on the TACoS dataset.
Therefore, we choose the CMCC loss as the final loss of our cross-modal cycle consistency.

\noindent \textbf{Analysis of the hyperparameters.}
To achieve the best performance, we analyze the impact of three hyperparameters: $\lambda_{\ell_{1}}$, $\lambda_{{IoU}}$ and $\lambda_f$.
Table \ref{tab:chaocan} shows the experimental results.
It can be observed that, with the increase of $\lambda_{\ell_{1}}$, $\lambda_{{IoU}}$ and $\lambda_f$, their performance follows a general trend, \textit{i.e.}, rises at first and then starts to decline. The optimal values of $\lambda_{\ell_{1}}$, $\lambda_{{IoU}}$ and $\lambda_f$ are 0.8, 0.5, and 0.2 respectively, where all the hyperparameters obtain the best performance.
Therefore, in our paper, we set $\lambda_{\ell_{1}}=0.8$, $\lambda_{{IoU}}=0.5$ and $\lambda_f=0.2$.

\noindent \textbf{Performance of different datasets.} To analyze the generalization ability of our HLGT, we test its running speed on different datasets. Table \ref{tab:three_data_performance} shows its performance on three datasets. On the one hand, although these datasets are under different scales, our HLGT has a similar  parameter scale for differnt datasets, which shows that HLGT can deal with different types of datasets with little or no model changes.
On the other hand, on the ActivityNet Captions and TACoS datasets, HLGT has the similar running speed (VPS). On the ActivityNet dataset, our HLGT deal with 190.76 videos per second, while it processes  196.54 videos on the TACoS dataset per second. It is because ActivityNet Captions and TACoS have similar average video length. The Charades-STA dataset has shorter video length, which leads to larger VPS.

\begin{figure}[t!]
\centering
\includegraphics[width=0.5\textwidth]{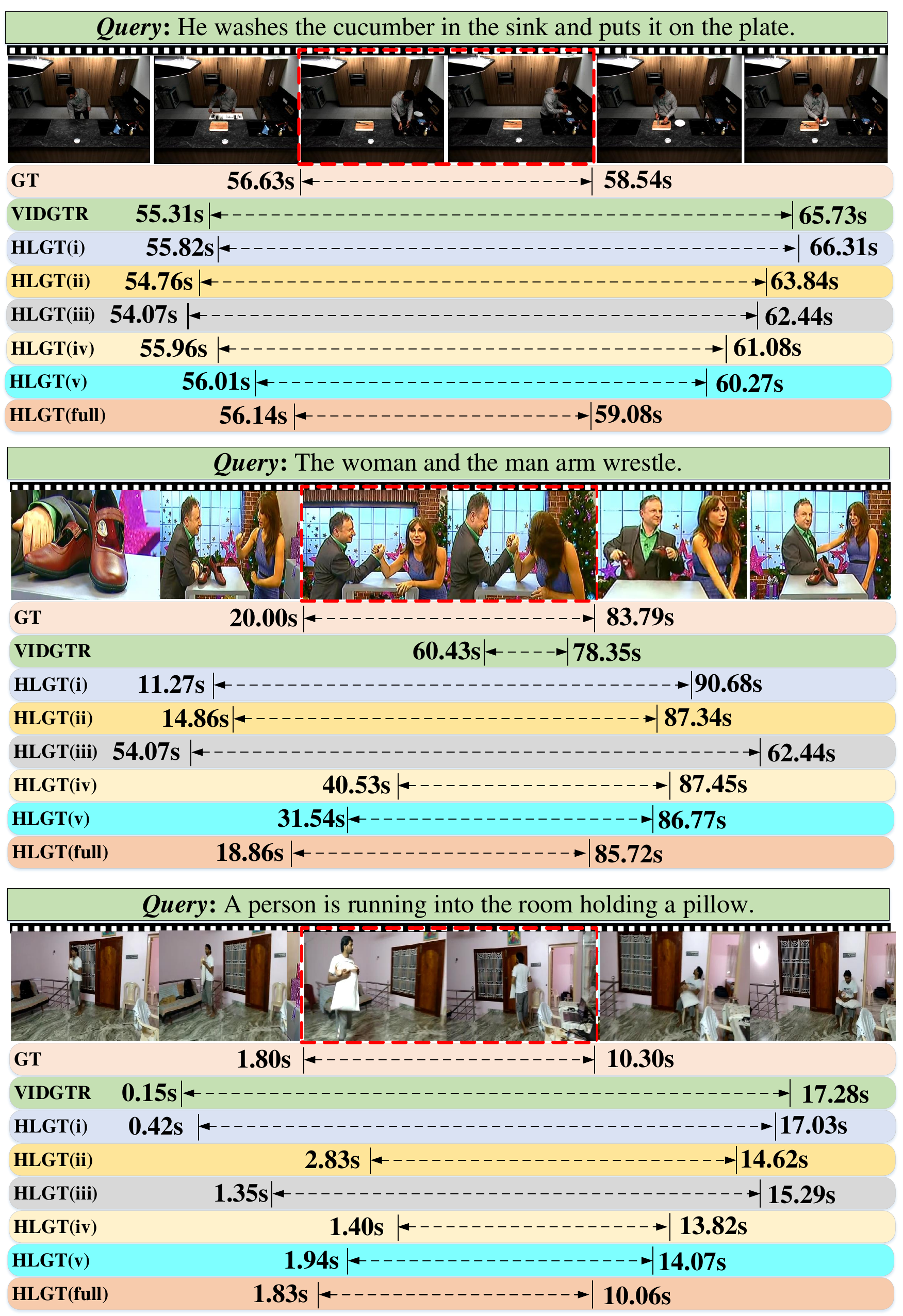}
\caption{Qualitative results sampled from all three datasets (top: TACoS, middle: ActivityNet Captions, bottom: Charades-STA).}
\label{fig:qualitative}
\end{figure}

\subsection{Visualization}
To investigate the grounding results of our HLGT, we provide two qualitative examples of HLGT and VIDGTR in Fig.~\ref{fig:qualitative}.  We can observe that HLGT achieves more precise localization than the state-of-the-art method VIDGTR.
The main reason is that VIDGTR only focuses on the frame-level visual features for encoding and ignores higher-level features (video- and clip-level features), which fails to capture the subtle cross-modal multi-granularity details and understand the complicated background visual content. Different from VIDGTR, HLGT can learn multi-level cross-modal interaction (clip-phrase and video-query interaction), thus capturing more fine-grained visual contexts for more accurate grounding.

\section{Conclusion}
In this paper, we proposed a novel Hierarchical Local-Global Transformer (HLGT) for temporal sentence grounding, which leverage this hierarchy information and model the interactions between multiple levels of granularity and different modalities for learning more fine-grained multi-modal representations. Experimental results on three challenging datasets (ActivityNet Captions, Charades-STA and TACoS) validate the effectiveness of our HLGT. In the future, we will apply HLGT to other tasks/datasets \cite{li2020hero,lei2020tvr} to further improve its generalization. Meanwhile, we also will introduce HLGT in the weakly-supervised manner to explore how to use more unannotated data in supervised manner.
\bibliographystyle{IEEEtran}
\bibliography{sample-base}

@inproceedings{zhang2021multi,
  title={Multi-stage aggregated transformer network for temporal language localization in videos},
  author={Zhang, Mingxing and Yang, Yang and Chen, Xinghan and Ji, Yanli and Xu, Xing and Li, Jingjing and Shen, Heng Tao},
  booktitle={CVPR},
  pages={12669--12678},
  year={2021}
}

@article{wang2022rcl,
  title={RCL: Recurrent Continuous Localization for Temporal Action Detection},
  author={Wang, Qiang and Zhang, Yanhao and Zheng, Yun and Pan, Pan},
  journal={arXiv preprint arXiv:2203.07112},
  year={2022}
}

@inproceedings{wang2021cycle,
  title={Cycle-consistent inverse gan for text-to-image synthesis},
  author={Wang, Hao and Lin, Guosheng and Hoi, Steven CH and Miao, Chunyan},
  booktitle={ACM MM},
  pages={630--638},
  year={2021}
}

@article{li2021referring,
  title={Referring transformer: A one-step approach to multi-task visual grounding},
  author={Li, Muchen and Sigal, Leonid},
  journal={NeurIPS},
  volume={34},
  year={2021}
}

@inproceedings{wang2021exploring,
  title={Exploring Sequence Feature Alignment for Domain Adaptive Detection Transformers},
  author={Wang, Wen and Cao, Yang and Zhang, Jing and He, Fengxiang and Zha, Zheng-Jun and Wen, Yonggang and Tao, Dacheng},
  booktitle={ACM MM},
  pages={1730--1738},
  year={2021}
}

@inproceedings{cao2021pursuit,
  title={On Pursuit of Designing Multi-modal Transformer for Video Grounding},
  author={Cao, Meng and Chen, Long and Shou, Mike Zheng and Zhang, Can and Zou, Yuexian},
  booktitle={EMNLP},
  pages={9810--9823},
  year={2021}
}

@inproceedings{wang2019learning,
  title={Learning correspondence from the cycle-consistency of time},
  author={Wang, Xiaolong and Jabri, Allan and Efros, Alexei A},
  booktitle={CVPR},
  pages={2566--2576},
  year={2019}
}

@inproceedings{lin2020weakly,
  title={Weakly-supervised video moment retrieval via semantic completion network},
  author={Lin, Zhijie and Zhao, Zhou and Zhang, Zhu and Wang, Qi and Liu, Huasheng},
  booktitle={AAAI},
  volume={34},
  number={07},
  pages={11539--11546},
  year={2020}
}

@article{liu2019cyclematch,
  title={CycleMatch: A cycle-consistent embedding network for image-text matching},
  author={Liu, Yu and Guo, Yanming and Liu, Li and Bakker, Erwin M and Lew, Michael S},
  journal={PR},
  volume={93},
  pages={365--379},
  year={2019},
  publisher={Elsevier}
}

@article{wang2022counterfactual,
  title={Counterfactual Cycle-Consistent Learning for Instruction Following and Generation in Vision-Language Navigation},
  author={Wang, Hanqing and Liang, Wei and Shen, Jianbing and Van Gool, Luc and Wang, Wenguan},
  journal={arXiv preprint arXiv:2203.16586},
  year={2022}
}

@inproceedings{shah2019cycle,
  title={Cycle-consistency for robust visual question answering},
  author={Shah, Meet and Chen, Xinlei and Rohrbach, Marcus and Parikh, Devi},
  booktitle={CVPR},
  year={2019}
}

@article{ding2021decoupling,
  title={Decoupling Zero-Shot Semantic Segmentation},
  author={Ding, Jian and Xue, Nan and Xia, Gui-Song and Dai, Dengxin},
  journal={arXiv preprint arXiv:2112.07910},
  year={2021}
}

@inproceedings{liu2021adaptive,
  title={Adaptive Proposal Generation Network for Temporal Sentence Localization in Videos},
  author={Liu, Daizong and Qu, Xiaoye and Dong, Jianfeng and Zhou, Pan},
  booktitle={EMNLP},
  pages={9292--9301},
  year={2021}
}

@InProceedings{PenningtonSM14,
  author    = {Jeffrey Pennington and Richard Socher and Christopher D. Manning},
  booktitle = {EMNLP},
  title     = {Glove: Global Vectors for Word Representation},
  year      = {2014},
  pages     = {1532--1543},
}

@article{hendrycks2016gaussian,
  title={Gaussian error linear units (gelus)},
  author={Hendrycks, Dan and Gimpel, Kevin},
  journal={arXiv preprint arXiv:1606.08415},
  year={2016}
}

@inproceedings{he2016deep,
  title={Deep residual learning for image recognition},
  author={He, Kaiming and Zhang, Xiangyu and Ren, Shaoqing and Sun, Jian},
  booktitle={CVPR},
  pages={770--778},
  year={2016}
}

@article{tang2021frame,
  title={Frame-wise cross-modal matching for video moment retrieval},
  author={Tang, Haoyu and Zhu, Jihua and Liu, Meng and Gao, Zan and Cheng, Zhiyong},
  journal={IEEE TMM},
  volume={24},
  pages={1338--1349},
  year={2021},
  publisher={IEEE}
}

@article{zhang2020temporal,
  title={Temporal textual localization in video via adversarial bi-directional interaction networks},
  author={Zhang, Zijian and Zhao, Zhou and Zhang, Zhu and Lin, Zhijie and Wang, Qi and Hong, Richang},
  journal={IEEE TMM},
  volume={23},
  pages={3306--3317},
  year={2020},
  publisher={IEEE}
}

@article{wang2022cross,
  title={Cross-modal dynamic networks for video moment retrieval with text query},
  author={Wang, Gongmian and Xu, Xing and Shen, Fumin and Lu, Huimin and Ji, Yanli and Shen, Heng Tao},
  journal={IEEE TMM},
  volume={24},
  pages={1221--1232},
  year={2022},
  publisher={IEEE}
}

@article{woo2022explore,
  title={Explore and Match: End-to-End Video Grounding with Transformer},
  author={Woo, Sangmin and Park, Jinyoung and Koo, Inyong and Lee, Sumin and Jeong, Minki and Kim, Changick},
  journal={arxiv},
  year={2022}
}

@inproceedings{carion2020end,
  title={End-to-end object detection with transformers},
  author={Carion, Nicolas and Massa, Francisco and Synnaeve, Gabriel and Usunier, Nicolas and Kirillov, Alexander and Zagoruyko, Sergey},
  booktitle={ECCV},
  pages={213--229},
  year={2020},
  organization={Springer}
}

@inproceedings{qu2020fine,
  title={Fine-grained iterative attention network for temporal language localization in videos},
  author={Qu, Xiaoye and Tang, Pengwei and Zou, Zhikang and Cheng, Yu and Dong, Jianfeng and Zhou, Pan and Xu, Zichuan},
  booktitle={ACM MM},
  pages={4280--4288},
  year={2020}
}

@inproceedings{zhang2019cross,
  title={Cross-modal interaction networks for query-based moment retrieval in videos},
  author={Zhang, Zhu and Lin, Zhijie and Zhao, Zhou and Xiao, Zhenxin},
  booktitle={SIGIR},
  pages={655--664},
  year={2019}
}

@article{song2021enhancing,
  title={Enhancing neural machine translation with dual-side multimodal awareness},
  author={Song, Yuqing and Chen, Shizhe and Jin, Qin and Luo, Wei and Xie, Jun and Huang, Fei},
  journal={IEEE TMM},
  year={2021},
  publisher={IEEE}
}

@article{zhang2022unified,
  title={Unified Adaptive Relevance Distinguishable Attention Network for Image-Text Matching},
  author={Zhang, Kun and Mao, Zhendong and Liu, Anan and Zhang, Yongdong},
  journal={IEEE TMM},
  year={2022}
}

@article{chen2022boosting,
  title={Boosting Vision-and-Language Navigation with Direction Guiding and Backtracing},
  author={Chen, Jingwen and Luo, Jianjie and Pan, Yingwei and Li, Yehao and Yao, Ting and Chao, Hongyang and Mei, Tao},
  journal={ACM TOMM},
  year={2022}
}

@article{fang2021animc,
  title={ANIMC: A Soft Approach for Autoweighted Noisy and Incomplete Multiview Clustering},
  author={Fang, Xiang and Hu, Yuchong and Zhou, Pan and Wu, Dapeng},
  journal={IEEE TAI},
  volume={3},
  number={2},
  pages={192--206},
  year={2021}
}

@article{fang2020v,
  title={V$^3$H: View variation and view heredity for incomplete multiview clustering},
  author={Fang, Xiang and Hu, Yuchong and Zhou, Pan and Wu, Dapeng Oliver},
  journal={IEEE TAI},
  volume={1},
  number={3},
  pages={233--247},
  year={2020}
}

@article{fang2021unbalanced,
  title={Unbalanced Incomplete Multi-view Clustering via the Scheme of View Evolution: Weak Views are Meat; Strong Views do Eat},
  author={Fang, Xiang and Hu, Yuchong and Zhou, Pan and Wu, Dapeng Oliver},
  journal={IEEE TETCI},
  year={2021}
}

@article{ma2022visualizing,
  title={Visualizing and Understanding Patch Interactions in Vision Transformer},
  author={Ma, Jie and Bai, Yalong and Zhong, Bineng and Zhang, Wei and Yao, Ting and Mei, Tao},
  journal={arXiv preprint arXiv:2203.05922},
  year={2022}
}

@article{li2022contextual,
  title={Contextual transformer networks for visual recognition},
  author={Li, Yehao and Yao, Ting and Pan, Yingwei and Mei, Tao},
  journal={IEEE TPAMI},
  year={2022}
}

@article{yao2022wave,
  title={Wave-ViT: Unifying Wavelet and Transformers for Visual Representation Learning},
  author={Yao, Ting and Pan, Yingwei and Li, Yehao and Ngo, Chong-Wah and Mei, Tao},
  journal={arXiv preprint arXiv:2207.04978},
  year={2022}
}

@article{yao2022dual,
  title={Dual Vision Transformer},
  author={Yao, Ting and Li, Yehao and Pan, Yingwei and Wang, Yu and Zhang, Xiao-Ping and Mei, Tao},
  journal={arXiv preprint arXiv:2207.04976},
  year={2022}
}

@article{wu2022blind,
  title={Blind Image Restoration Based on Cycle-Consistent Network},
  author={Wu, Shixiang and Dong, Chao and Qiao, Yu},
  journal={IEEE TMM},
  year={2022}
}

@inproceedings{caba2015activitynet,
  title={Activitynet: A large-scale video benchmark for human activity understanding},
  author={Caba Heilbron, Fabian and Escorcia, Victor and Ghanem, Bernard and Carlos Niebles, Juan},
  booktitle={CVPR},
  pages={961--970},
  year={2015}
}

@inproceedings{krishna2017dense,
  title={Dense-captioning events in videos},
  author={Krishna, Ranjay and Hata, Kenji and Ren, Frederic and Fei-Fei, Li and Carlos Niebles, Juan},
  booktitle={ICCV},
  pages={706--715},
  year={2017}
}

@article{regneri2013grounding,
  title={Grounding action descriptions in videos},
  author={Regneri, Michaela and Rohrbach, Marcus and Wetzel, Dominikus and Thater, Stefan and Schiele, Bernt and Pinkal, Manfred},
  journal={TACL},
  volume={1},
  pages={25--36},
  year={2013}
}

@inproceedings{gao2017tall,
  title={Tall: Temporal activity localization via language query},
  author={Gao, Jiyang and Sun, Chen and Yang, Zhenheng and Nevatia, Ram},
  booktitle={ICCV},
  pages={5267--5275},
  year={2017}
}

@inproceedings{anne2017localizing,
  title={Localizing moments in video with natural language},
  author={Anne Hendricks, Lisa and Wang, Oliver and Shechtman, Eli and Sivic, Josef and Darrell, Trevor and Russell, Bryan},
  booktitle={ICCV},
  year={2017}
}

@inproceedings{yuan2019find,
  title={To find where you talk: Temporal sentence localization in video with attention based location regression},
  author={Yuan, Yitian and Mei, Tao and Zhu, Wenwu},
  booktitle={AAAI},
  volume={33},
  pages={9159--9166},
  year={2019}
}

@inproceedings{liu2018attentive,
  title={Attentive moment retrieval in videos},
  author={Liu, Meng and Wang, Xiang and Nie, Liqiang and He, Xiangnan and Chen, Baoquan and Chua, Tat-Seng},
  booktitle={SIGIR},
  pages={15--24},
  year={2018}
}

@inproceedings{chen2019semantic,
  title={Semantic proposal for activity localization in videos via sentence query},
  author={Chen, Shaoxiang and Jiang, Yu-Gang},
  booktitle={AAAI},
  volume={33},
  pages={8199--8206},
  year={2019}
}

@inproceedings{chen2018temporally,
  title={Temporally grounding natural sentence in video},
  author={Chen, Jingyuan and Chen, Xinpeng and Ma, Lin and Jie, Zequn and Chua, Tat-Seng},
  booktitle={EMNLP},
  pages={162--171},
  year={2018}
}

@inproceedings{wang2019temporally,
  title={Temporally Grounding Language Queries in Videos by Contextual Boundary-aware Prediction},
  author={Wang, Jingwen and Ma, Lin and Jiang, Wenhao},
  booktitle={AAAI},
  year={2020}
}

@inproceedings{chenrethinking,
  title={Rethinking the Bottom-Up Framework for Query-based Video Localization},
  author={Chen, Long and Lu, Chujie and Tang, Siliang and Xiao, Jun and Zhang, Dong and Tan, Chilie and Li, Xiaolin},
  booktitle={AAAI},
  year={2020}
}

@inproceedings{hu2016natural,
  title={Natural language object retrieval},
  author={Hu, Ronghang and Xu, Huazhe and Rohrbach, Marcus and Feng, Jiashi and Saenko, Kate and Darrell, Trevor},
  booktitle={CVPR},
  pages={4555--4564},
  year={2016}
}

@inproceedings{sigurdsson2016hollywood,
  title={Hollywood in homes: Crowdsourcing data collection for activity understanding},
  author={Sigurdsson, Gunnar A and Varol, G{\"u}l and Wang, Xiaolong and Farhadi, Ali and Laptev, Ivan and Gupta, Abhinav},
  booktitle={ECCV},
  pages={510--526},
  year={2016}
}

@inproceedings{zhang2019learning,
  title={Learning 2D Temporal Adjacent Networks for Moment Localization with Natural Language},
  author={Zhang, Songyang and Peng, Houwen and Fu, Jianlong and Luo, Jiebo},
  booktitle={AAAI},
  year={2020}
}

@inproceedings{mun2020local,
  title={Local-Global Video-Text Interactions for Temporal Grounding},
  author={Mun, Jonghwan and Cho, Minsu and Han, Bohyung},
  booktitle={CVPR},
  pages={10810--10819},
  year={2020}
}

@inproceedings{zeng2020dense,
  title={Dense regression network for video grounding},
  author={Zeng, Runhao and Xu, Haoming and Huang, Wenbing and Chen, Peihao and Tan, Mingkui and Gan, Chuang},
  booktitle={CVPR},
  pages={10287--10296},
  year={2020}
}

@inproceedings{liu2020jointly,
  title={Jointly Cross-and Self-Modal Graph Attention Network for Query-Based Moment Localization},
  author={Liu, Daizong and Qu, Xiaoye and Liu, Xiao-Yang and Dong, Jianfeng and Zhou, Pan and Xu, Zichuan},
  booktitle={ACM MM},
  pages={4070--4078},
  year={2020}
}

@inproceedings{li2020hero,
  title={HERO: Hierarchical Encoder for Video+ Language Omni-representation Pre-training},
  author={Li, Linjie and Chen, Yen-Chun and Cheng, Yu and Gan, Zhe and Yu, Licheng and Liu, Jingjing},
  booktitle={EMNLP},
  year={2020}
}

@inproceedings{liu2021context,
  title={Context-aware Biaffine Localizing Network for Temporal Sentence Grounding},
  author={Liu, Daizong and Qu, Xiaoye and Dong, Jianfeng and Zhou, Pan and Cheng, Yu and Wei, Wei and Xu, Zichuan and Xie, Yulai},
  booktitle={CVPR},
  year={2021}
}

@inproceedings{xiao2021boundary,
  title={Boundary Proposal Network for Two-Stage Natural Language Video Localization},
  author={Xiao, Shaoning and Chen, Long and Zhang, Songyang and Ji, Wei and Shao, Jian and Ye, Lu and Xiao, Jun},
  booktitle={AAAI},
  year={2021}
}

@inproceedings{lei2020tvr,
  title={TVR: A Large-Scale Dataset for Video-Subtitle Moment Retrieval},
  author={Lei, Jie and Yu, Licheng and Berg, Tamara L and Bansal, Mohit},
  booktitle={ECCV},
  year={2020}
}

@inproceedings{he2021end,
  title={End-to-End Video Object Detection with Spatial-Temporal Transformers},
  author={He, Lu and Zhou, Qianyu and Li, Xiangtai and Niu, Li and Cheng, Guangliang and Li, Xiao and Liu, Wenxuan and Tong, Yunhai and Ma, Lizhuang and Zhang, Liqing},
  booktitle={ACM MM},
  pages={1507--1516},
  year={2021}
}

@article{ging2020coot,
  title={Coot: Cooperative hierarchical transformer for video-text representation learning},
  author={Ging, Simon and Zolfaghari, Mohammadreza and Pirsiavash, Hamed and Brox, Thomas},
  journal={NeurIPS},
  volume={33},
  pages={22605--22618},
  year={2020}
}

@inproceedings{zhang2018cross,
  title={Cross-modal and hierarchical modeling of video and text},
  author={Zhang, Bowen and Hu, Hexiang and Sha, Fei},
  booktitle={ECCV},
  pages={374--390},
  year={2018}
}

@inproceedings{zhang2020span,
  title={Span-based Localizing Network for Natural Language Video Localization},
  author={Zhang, Hao and Sun, Aixin and Jing, Wei and Zhou, Joey Tianyi},
  booktitle={ACL},
  pages={6543--6554},
  year={2020}
}

@inproceedings{liu2022unsupervised,
  title={Unsupervised Temporal Video Grounding with Deep Semantic Clustering},
  author={Liu, Daizong and Qu, Xiaoye and Wang, Yinzhen and Di, Xing and Zou, Kai and Cheng, Yu and Xu, Zichuan and Zhou, Pan},
  booktitle={AAAI},
  year={2022}
}

@inproceedings{nan2021interventional,
  title={Interventional Video Grounding with Dual Contrastive Learning},
  author={Nan, Guoshun and Qiao, Rui and Xiao, Yao and Liu, Jun and Leng, Sicong and Zhang, Hao and Lu, Wei},
  booktitle={CVPR},
  year={2021}
}

@article{kingma2014adam,
  title={Adam: A method for stochastic optimization},
  author={Kingma, Diederik P and Ba, Jimmy},
  journal={arXiv preprint arXiv:1412.6980},
  year={2014}
}

@inproceedings{dosovitskiy2020image,
  title={An Image is Worth 16x16 Words: Transformers for Image Recognition at Scale},
  author={Dosovitskiy, Alexey and Beyer, Lucas and Kolesnikov, Alexander and Weissenborn, Dirk and Zhai, Xiaohua and Unterthiner, Thomas and Dehghani, Mostafa and Minderer, Matthias and Heigold, Georg and Gelly, Sylvain and others},
  booktitle={ICLR},
  year={2020}
}

@inproceedings{min2020isia,
  title={Isia food-500: A dataset for large-scale food recognition via stacked global-local attention network},
  author={Min, Weiqing and Liu, Linhu and Wang, Zhiling and Luo, Zhengdong and Wei, Xiaoming and Wei, Xiaolin and Jiang, Shuqiang},
  booktitle={ACM MM},
  pages={393--401},
  year={2020}
}

@article{zhang2021natural,
  title={Natural language video localization: A revisit in span-based question answering framework},
  author={Zhang, Hao and Sun, Aixin and Jing, Wei and Zhen, Liangli and Zhou, Joey Tianyi and Goh, Rick Siow Mong},
  journal={IEEE TPAMI},
  year={2021},
}

@article{xu2022mdan,
  title={MDAN: Multi-level Dependent Attention Network for Visual Emotion Analysis},
  author={Xu, Liwen and Wang, Zhengtao and Wu, Bin and Lui, Simon},
  journal={arXiv preprint arXiv:2203.13443},
  year={2022}
}

@article{zhang2022vitaev2,
  title={ViTAEv2: Vision Transformer Advanced by Exploring Inductive Bias for Image Recognition and Beyond},
  author={Zhang, Qiming and Xu, Yufei and Zhang, Jing and Tao, Dacheng},
  journal={arXiv preprint arXiv:2202.10108},
  year={2022}
}

@inproceedings{zhang2021token,
  title={Token shift transformer for video classification},
  author={Zhang, Hao and Hao, Yanbin and Ngo, Chong-Wah},
  booktitle={ACM MM},
  pages={917--925},
  year={2021}
}

@inproceedings{hou2021multiview,
  title={Multiview detection with shadow transformer (and view-coherent data augmentation)},
  author={Hou, Yunzhong and Zheng, Liang},
  booktitle={ACM MM},
  pages={1673--1682},
  year={2021}
}

@inproceedings{vaswani2017attention,
  title={Attention is all you need},
  author={Vaswani, Ashish and Shazeer, Noam and Parmar, Niki and Uszkoreit, Jakob and Jones, Llion and Gomez, Aidan N and Kaiser, {\L}ukasz and Polosukhin, Illia},
  booktitle={NIPS},
  pages={5998--6008},
  year={2017}
}

@inproceedings{xu2019multilevel,
  title={Multilevel language and vision integration for text-to-clip retrieval},
  author={Xu, Huijuan and He, Kun and Plummer, Bryan A and Sigal, Leonid and Sclaroff, Stan and Saenko, Kate},
  booktitle={AAAI },
  volume={33},
  pages={9062--9069},
  year={2019}
}

@article{liu2022exploring,
  title={Exploring Optical-Flow-Guided Motion and Detection-Based Appearance for Temporal Sentence Grounding},
  author={Liu, Daizong and Fang, Xiang and Hu, Wei and Zhou, Pan},
  journal={arXiv preprint arXiv:2203.02966},
  year={2022}
}

@InProceedings{yuan2019semantic,
  author    = {Yitian Yuan and Lin Ma and Jingwen Wang and Wei Liu and Wenwu Zhu},
  booktitle = {NeurIPS},
  title     = {Semantic Conditioned Dynamic Modulation for Temporal Sentence Grounding in Videos},
  year      = {2019},
}

@inproceedings{liu2022memory,
  title={Memory-Guided Semantic Learning Network for Temporal Sentence Grounding},
  author={Liu, Daizong and Qu, Xiaoye and Di, Xing and Cheng, Yu and Xu, Zichuan Xu and Zhou, Pan},
  booktitle={AAAI},
  year={2022}
}

@article{fang2025your,
  title={Your data is not perfect: Towards cross-domain out-of-distribution detection in class-imbalanced data},
  author={Fang, Xiang and Easwaran, Arvind and Genest, Blaise and Suganthan, Ponnuthurai Nagaratnam},
  journal={Expert Systems with Applications},
  year={2025}
}

@article{fang2022multi,
  title={Multi-modal cross-domain alignment network for video moment retrieval},
  author={Fang, Xiang and Liu, Daizong and Zhou, Pan and Hu, Yuchong},
  journal={IEEE Transactions on Multimedia},
  volume={25},
  pages={7517--7532},
  year={2022},
  publisher={IEEE}
}

@inproceedings{fang2026cogniVerse,
  title={CogniVerse: Revolutionizing Multi-modal Retrieval-Augmented Generation with Cognitive Reflection and Geometric Reasoning},
  author={Fang, Xiang and Fang, Wanlong and Wang, Changshuo},
  booktitle={Proceedings of the IEEE/CVF Conference on Computer Vision and Pattern Recognition},
  year={2026}
}

@inproceedings{fang2023you,
  title={You can ground earlier than see: An effective and efficient pipeline for temporal sentence grounding in compressed videos},
  author={Fang, Xiang and Liu, Daizong and Zhou, Pan and Nan, Guoshun},
  booktitle={Proceedings of the IEEE/CVF Conference on Computer Vision and Pattern Recognition},
  pages={2448--2460},
  year={2023}
}

@inproceedings{fang2025hierarchical,
  title={Hierarchical Semantic-Augmented Navigation: Optimal Transport and Graph-Driven Reasoning for Vision-Language Navigation},
  author={Fang, Xiang and Fang, Wanlong and Wang, Changshuo},
  booktitle={Advances in Neural Information Processing Systems},
  year={2025}
}

@inproceedings{fang2025adaptive,
  title={Adaptive Multi-prompt Contrastive Network for Few-shot Out-of-distribution Detection},
  author={Fang, Xiang and Easwaran, Arvind and Genest, Blaise},
  booktitle={International Conference on Machine Learning},
  year={2025}
}

@inproceedings{fang2026slap,
  title={SLAP: The Semantic Least Action Principle for Variational Video-Language Modeling},
  author={Fang, Xiang and Fang, Wanlong},
  booktitle={International Conference on Machine Learning},
  year={2026}
}

@inproceedings{fang2026immuno,
  title={Immuno-VLM: Immunizing Large Vision-Language Models via Generative Semantic Antibodies for Open-World Trustworthiness},
  author={Fang, Xiang and Fang, Wanlong and Ji, Wei},
  booktitle={International Conference on Machine Learning},
  year={2026}
}

@inproceedings{fang2026disentangling,
  title={Disentangling Adversarial Prompts: A Semantic-Graph Defense for Robust LLM Security},
  author={Fang, Xiang and Fang, Wanlong},
 booktitle={Proceedings of the AAAI Conference on Artificial Intelligence},
year={2026}
}

@inproceedings{fang2026advancing,
  title={Advancing Out-of-Distribution Detection Across Diverse Scenarios},
  author={Fang, Xiang},
  booktitle={Proceedings of the AAAI Conference on Artificial Intelligence},
  volume={40},
  number={48},
  pages={41042--41043},
  year={2026}
}

@inproceedings{fang2026unveiling,
  title={Unveiling the Fragility of Vision-Language Models: Multi-Modal Adversarial Synergy via Texture-Constrained Perturbations and Cross-Modal Optimization},
  author={Fang, Xiang and Fang, Wanlong and Wang, Changshuo},
 booktitle={Proceedings of the AAAI Conference on Artificial Intelligence},
year={2026}
}

@inproceedings{fang2026rethinking,
  title={Rethinking Video-language Model From the Language Input Perspective},
  author={Fang, Xiang and Fang, Wanlong and Wang, Changshuo and Qu, Xiaoye and Liu, Daizong},
 booktitle={Proceedings of the AAAI Conference on Artificial Intelligence},
year={2026}
}

@inproceedings{fang2026towards,
  title={Towards Unified Vision-Language Models With Incomplete Multi-Modal Inputs},
  author={Fang, Xiang and Fang, Wanlong and Wang, Changshuo and Tang, Keke and Liu, Daizong and Wang, Siyi and Ji, Wei},
 booktitle={Proceedings of the AAAI Conference on Artificial Intelligence},
year={2026}
}

@inproceedings{fang2025multi,
  title={Multi-pair temporal sentence grounding via multi-thread knowledge transfer network},
  author={Fang, Xiang and Fang, Wanlong and Wang, Changshuo and Liu, Daizong and Tang, Keke and Dong, Jianfeng and Zhou, Pan and Li, Beibei},
  booktitle={Proceedings of the AAAI Conference on Artificial Intelligence},
  volume={39},
  number={3},
  pages={2915--2923},
  year={2025}
}

@inproceedings{fang2024fewer,
  title={Fewer Steps, Better Performance: Efficient Cross-Modal Clip Trimming for Video Moment Retrieval Using Language},
  author={Fang, Xiang and Liu, Daizong and Fang, Wanlong and Zhou, Pan and Xu, Zichuan and Xu, Wenzheng and Chen, Junyang and Li, Renfu},
  booktitle={Proceedings of the AAAI Conference on Artificial Intelligence},
  volume={38},
  number={2},
  pages={1735--1743},
  year={2024}
}

@inproceedings{fang2024multi,
  title={Multi-Pair Temporal Sentence Grounding via Multi-Thread Knowledge Transfer Network},
  author={Fang, Xiang and Fang, Wanlong and Wang, Changshuo and Liu, Daizong and Tang, Keke and Dong, Jianfeng and Zhou, Pan and Li, Beibei},
  booktitle={Proceedings of the AAAI Conference on Artificial Intelligence},
  year={2025}
}

@inproceedings{fang2025turing,
  title={Turing Patterns for Multimedia: Reaction-Diffusion Multi-Modal Fusion for Language-Guided Video Moment Retrieval},
  author={Fang, Xiang and Fang, Wanlong and Ji, Wei and Chua, Tat-Seng},
  booktitle={ACM International Conference on Multimedia},
  year={2025}
}

@inproceedings{fang2024not,
  title={Not all inputs are valid: Towards open-set video moment retrieval using language},
  author={Fang, Xiang and Fang, Wanlong and Liu, Daizong and Qu, Xiaoye and Dong, Jianfeng and Zhou, Pan and Li, Renfu and Xu, Zichuan and Chen, Lixing and Zheng, Panpan and others},
  booktitle={Proceedings of the 32nd ACM International Conference on Multimedia},
  pages={28--37},
  year={2024}
}

@inproceedings{fang2024rethinking,
  title={Rethinking Weakly-supervised Video Temporal Grounding From a Game Perspective},
  author={Fang, Xiang and Xiong, Zeyu and Fang, Wanlong and Qu, Xiaoye and Chen, Chen and Dong, Jianfeng and Tang, Keke and Zhou, Pan and Cheng, Yu and Liu, Daizong},
  booktitle={European Conference on Computer Vision},
  year={2024},
  organization={Springer}
}

@inproceedings{fang2023annotations,
  title={Annotations Are Not All You Need: A Cross-modal Knowledge Transfer Network for Unsupervised Temporal Sentence Grounding},
  author={Fang, Xiang and Liu, Daizong and Fang, Wanlong and Zhou, Pan and Cheng, Yu and Tang, Keke and Zou, Kai},
  booktitle={Findings of the Association for Computational Linguistics: EMNLP 2023},
  pages={8721--8733},
  year={2023}
}

@article{fang2025adaptivetai,
  title={Adaptive Hierarchical Graph Cut for Multi-granularity Out-of-distribution Detection},
  author={Fang, Xiang and Easwaran, Arvind and Genest, Blaise and Suganthan, Ponnuthurai Nagaratnam},
  journal={IEEE Transactions on Artificial Intelligence},
  year={2025}
}

@article{fang2020double,
  title={Double self-weighted multi-view clustering via adaptive view fusion},
  author={Fang, Xiang and Hu, Yuchong},
  journal={arXiv preprint arXiv:2011.10396},
  year={2020}
}

@article{liu2023exploring,
  title={Exploring optical-flow-guided motion and detection-based appearance for temporal sentence grounding},
  author={Liu, Daizong and Fang, Xiang and Hu, Wei and Zhou, Pan},
  journal={IEEE Transactions on Multimedia},
  volume={25},
  pages={8539--8553},
  year={2023},
  publisher={IEEE}
}

@inproceedings{wang2025taylor,
  title={Taylor series-inspired local structure fitting network for few-shot point cloud semantic segmentation},
  author={Wang, Changshuo and He, Shuting and Fang, Xiang and Wu, Meiqing and Lam, Siew-Kei and Tiwari, Prayag},
  booktitle={Proceedings of the AAAI Conference on Artificial Intelligence},
  volume={39},
  number={7},
  pages={7527--7535},
  year={2025}
}

@inproceedings{wang2025point,
  title={Point clouds meets physics: Dynamic acoustic field fitting network for point cloud understanding},
  author={Wang, Changshuo and He, Shuting and Fang, Xiang and Han, Jiawei and Liu, Zhonghang and Ning, Xin and Li, Weijun and Tiwari, Prayag},
  booktitle={Proceedings of the Computer Vision and Pattern Recognition Conference},
  pages={22182--22192},
  year={2025}
}

@inproceedings{wang2025dypolyseg,
  title={DyPolySeg: Taylor Series-Inspired Dynamic Polynomial Fitting Network for Few-shot Point Cloud Semantic Segmentation},
  author={Wang, Changshuo and Fang, Xiang and Tiwari, Prayag},
  booktitle={Forty-second International Conference on Machine Learning},
  year={2025}
}

@article{wang2026reasoning,
  title={Reasoning beyond points: A visual introspective approach for few-shot 3d segmentation},
  author={Wang, Changshuo and He, Shuting and Fang, Xiang and Hu, Zhijian and Huang, Jia-Hong and Shen, Yixian and Tiwari, Prayag},
  journal={Advances in Neural Information Processing Systems},
  volume={38},
  pages={117394--117414},
  year={2026}
}

@article{wang2026from,
  title={From Coarse to Fine: Deep Prototype Refinement Network for Few-Shot Point Cloud Semantic Segmentation},
  author={Wang, Changshuo and He, Shuting and Fang, Xiang and Li, Weijun and Gao, Xingyu and Liu, Zhonghang and Tiwari, Prayag and Kanoulas, Dimitrios},
  journal={International Conference on Machine Learning},
  year={2026}
}

@article{wang2026topadapter,
  title={TopAdapter: Topology-Aware Prompt Tuning for Efficient Point Cloud Understanding},
  author={Wang, Changshuo and He, Shuting and Fang, Xiang and Li, Weijun and Shen, Yixian and Xu, Mingkun and Sun, Zhongtian and Tiwari, Prayag},
  journal={International Conference on Machine Learning},
  year={2026}
}

@inproceedings{wang2026biologically,
  title={Biologically-Inspired Evolutionary Domain Symbiosis for Few-shot and Zero-shot Point Cloud Semantic Segmentation},
  author={Wang, Changshuo and Hu, Zhijian and Fang, Xiang and Yu, Zai Yang and Wu, Yibin and Xu, Mingkun and Wang, Yusong and Gao, Xingyu and Tiwari, Prayag},
  booktitle={Proceedings of the AAAI Conference on Artificial Intelligence},
  volume={40},
  number={12},
  pages={9666--9674},
  year={2026}
}

@inproceedings{yang2025eood,
  title={EOOD: Entropy-based Out-of-distribution Detection},
  author={Yang, Guide and Hou, Chao and Peng, Weilong and Fang, Xiang and Nie, Yongwei and Zhu, Peican and Tang, Keke},
  booktitle={2025 International Joint Conference on Neural Networks (IJCNN)},
  pages={1--8},
  year={2025},
  organization={IEEE}
}

@inproceedings{wang2025reducing
,
  title={Reducing T-Depth and T-Count in Quantum Multiplication Using Compressor Primitives},
  author={Wang, Siyi and Dutta, Suman and Lee, Wei Jie Bryan and Feng, Jerrie and Fang, Xiang and Chattopadhyay, Anupam},
  booktitle={Proceedings of the Great Lakes Symposium on VLSI 2025},
  pages={35--40},
  year={2025}
}

@inproceedings{lei2025exploring,
  title={Exploring Disentangled Appearance-Motion Contexts for Temporal Activity Localization},
  author={Lei, Huashuo and Cai, Xiaowen and Liu, Daizong and Fang, Xiang and Qu, Xiaoye and Dong, Jianfeng and Yu, Jixiang and Jin, Keyan},
  booktitle={2025 International Joint Conference on Neural Networks (IJCNN)},
  pages={1--8},
  year={2025},
  organization={IEEE}
}

@inproceedings{zhang2025monoattack,
  title={MonoAttack: A Strong Attack Framework with Depth-Migration and Attribute-Tampering for Monocular 3D Object Detection},
  author={Zhang, Xiayue and Lei, Huashuo and Liu, Daizong and Qu, Xiaoye and Fang, Xiang and Guan, Runwei and Jin, Keyan},
  booktitle={2025 International Joint Conference on Neural Networks (IJCNN)},
  pages={1--8},
  year={2025},
  organization={IEEE}
}

@inproceedings{zhang2025manipulating,
  title={Manipulating the Bounding Box: Multimodal Controlled Backdoor Attacks on 3D Visual Grounding Models},
  author={Zhang, Xiayue and Lei, Huashuo and Liu, Daizong and Qu, Xiaoye and Fang, Xiang and Guan, Runwei and Jin, Keyan},
  booktitle={2025 International Joint Conference on Neural Networks (IJCNN)},
  pages={1--8},
  year={2025},
  organization={IEEE}
}

@article{wang2025prototype,
  title={Prototype-driven structure synergy network for remote sensing images segmentation},
  author={Wang, Junyi and Li, Jinjiang and Fan, Guodong and Ju, Yakun and Fang, Xiang and Kot, Alex C},
  journal={IEEE Transactions on Geoscience and Remote Sensing},
  year={2025},
  publisher={IEEE}
}

@inproceedings{wang2025seeing,
  title={Seeing the Overlooked: Bio-Visual Inspired Weak Saliency Feedback Transformer for Person Re-identification},
  author={Wang, Changshuo and He, Shuting and Fang, Xiang and Nan, Fangzhe and Tiwari, Prayag},
  booktitle={Proceedings of the 33rd ACM International Conference on Multimedia},
  pages={3192--3201},
  year={2025}
}

@inproceedings{fang2026align,
  title={To align or not to align: Strategic multimodal representation alignment for optimal performance},
  author={Fang, Wanlong and Zhang, Tianle and Chan, Alvin},
  booktitle={Proceedings of the AAAI Conference on Artificial Intelligence},
  volume={40},
  number={25},
  pages={21056--21064},
  year={2026}
}

@article{liu2023conditional,
  title={Conditional video diffusion network for fine-grained temporal sentence grounding},
  author={Liu, Daizong and Zhu, Jiahao and Fang, Xiang and Xiong, Zeyu and Wang, Huan and Li, Renfu and Zhou, Pan},
  journal={IEEE Transactions on Multimedia},
  volume={26},
  pages={5461--5476},
  year={2023},
  publisher={IEEE}
}

@article{liu2024pandora,
  title={Pandora's box: Towards building universal attackers against real-world large vision-language models},
  author={Liu, Daizong and Yang, Mingyu and Qu, Xiaoye and Zhou, Pan and Fang, Xiang and Tang, Keke and Wan, Yao and Sun, Lichao},
  journal={Advances in Neural Information Processing Systems},
  volume={37},
  pages={52127--52158},
  year={2024}
}

@inproceedings{liu2026attacking,
  title={Attacking Gray-Box Large Vision-Language Models with Adaptive SVD-Structured Adversarial Alignment},
  author={Liu, Daizong and Cai, Xiaowen and Dong, Junhao and Guo, Zhongliang and Qu, Xiaoye and Guan, Runwei and Fang, Xiang and Ye, Dengpan},
  booktitle={International Conference on Machine Learning},
  year={2026}
}

@inproceedings{liu2024unsupervised,
  title={Unsupervised domain adaptative temporal sentence localization with mutual information maximization},
  author={Liu, Daizong and Fang, Xiang and Qu, Xiaoye and Dong, Jianfeng and Yan, He and Yang, Yang and Zhou, Pan and Cheng, Yu},
  booktitle={Proceedings of the AAAI Conference on Artificial Intelligence},
  volume={38},
  number={4},
  pages={3567--3575},
  year={2024}
}

@inproceedings{liu2023hypotheses,
  title={Hypotheses tree building for one-shot temporal sentence localization},
  author={Liu, Daizong and Fang, Xiang and Zhou, Pan and Di, Xing and Lu, Weining and Cheng, Yu},
  booktitle={Proceedings of the AAAI Conference on Artificial Intelligence},
  volume={37},
  number={2},
  pages={1640--1648},
  year={2023}
}

@inproceedings{tang2024reparameterization,
  title={Reparameterization head for efficient multi-input networks},
  author={Tang, Keke and Zhao, Wenyu and Peng, Weilong and Fang, Xiang and Cui, Xiaodong and Zhu, Peican and Tian, Zhihong},
  booktitle={ICASSP 2024-2024 IEEE International Conference on Acoustics, Speech and Signal Processing (ICASSP)},
  pages={6190--6194},
  year={2024},
  organization={IEEE}
}

@article{xiong2024rethinking,
  title={Rethinking video sentence grounding from a tracking perspective with memory network and masked attention},
  author={Xiong, Zeyu and Liu, Daizong and Fang, Xiang and Qu, Xiaoye and Dong, Jianfeng and Zhu, Jiahao and Tang, Keke and Zhou, Pan},
  journal={IEEE Transactions on Multimedia},
  volume={26},
  pages={11204--11218},
  year={2024},
  publisher={IEEE}
}

@inproceedings{tang2025simplification,
  title={Simplification is all you need against out-of-distribution overconfidence},
  author={Tang, Keke and Hou, Chao and Peng, Weilong and Fang, Xiang and Wu, Zhize and Nie, Yongwei and Wang, Wenping and Tian, Zhihong},
  booktitle={Proceedings of the Computer Vision and Pattern Recognition Conference},
  pages={5030--5040},
  year={2025}
}

@article{cai2026towards,
  title={Towards building model/prompt-transferable attackers against large vision-language models},
  author={Cai, Xiaowen and Liu, Daizong and Qu, Xiaoye and Fang, Xiang and Dong, Jianfeng and Tang, Keke and Zhou, Pan and Sun, Lichao and Hu, Wei},
  journal={Advances in Neural Information Processing Systems},
  volume={38},
  pages={174022--174058},
  year={2026}
}

@article{yan2026fit,
  title={Fit the distribution: Cross-image/prompt adversarial attacks on multimodal large language models},
  author={Yan, Hai and Ma, Haijian and Cai, Xiaowen and Liu, Daizong and Yuan, Zenghui and Qu, Xiaoye and Dong, Jianfeng and Guan, Runwei and Fang, Xiang and He, Hongyang and others},
  journal={Advances in Neural Information Processing Systems},
  volume={38},
  pages={75204--75247},
  year={2026}
}

@inproceedings{liu2024towards,
  title={Towards robust temporal activity localization learning with noisy labels},
  author={Liu, Daizong and Qu, Xiaoye and Fang, Xiang and Dong, Jianfeng and Zhou, Pan and Nan, Guoshun and Tang, Keke and Fang, Wanlong and Cheng, Yu},
  booktitle={Proceedings of the 2024 Joint International Conference on Computational Linguistics, Language Resources and Evaluation (LREC-COLING 2024)},
  pages={16630--16642},
  year={2024}
}

@inproceedings{cai2025imperceptible,
  title={Imperceptible Beam-Sensitive Adversarial Attacks for LiDAR-based Object Detection in Autonomous Driving},
  author={Cai, Fuyao and Liu, Daizong and Fang, Xiang and Yu, Jixiang and Tang, Keke and Zhou, Pan},
  booktitle={2025 IEEE International Conference on Multimedia and Expo (ICME)},
  pages={1--6},
  year={2025},
  organization={IEEE}
}

@article{kuai2026dynamic,
  title={Dynamic Graph-enhanced Event Refinement for Temporal Sentence Grounding of Micro-moments},
  author={Kuai, Mingjin and Qin, You and Fang, Xiang and Ji, Wei and Zimmermann, Roger},
  journal={IEEE Transactions on Multimedia},
  year={2026},
  publisher={IEEE}
}

@inproceedings{fang2026towardsicml,
  title={Towards Understanding Modality Interaction in Multimodal Language Models via Partial Information Decomposition},
  author={Fang, Wanlong and Zhang, Tianle and Tao, Wen and Chan, Alvin},
  booktitle={International Conference on Machine Learning},
  year={2026}
}

\end{document}